\documentclass[aps,prb,twocolumn]{revtex4}

\usepackage{graphicx}
\usepackage{dcolumn}
\usepackage{bm}
\usepackage{amsmath,amssymb}
\begin{document}

\newcommand{\bk}{{\bf k}}
\newcommand{\bp}{{\bf p}}
\newcommand{\bv}{{\bf v}}
\newcommand{\bq}{{\bf q}}
\newcommand{\bt}{{\bf \tau}}
\newcommand{\bs}{{\bf s}}
\newcommand{\bmm}{{\bf m}}
\newcommand{\tbq}{\tilde{\bf q}}
\newcommand{\tq}{\tilde{q}}
\newcommand{\tPsi}{\tilde{\Psi}}
\newcommand{\bQ}{{\bf Q}}
\newcommand{\br}{{\bf r}}
\newcommand{\bR}{{\bf R}}
\newcommand{\bB}{{\bf B}}
\newcommand{\bE}{{\bf E}}
\newcommand{\bA}{{\bf A}}
\newcommand{\bK}{{\bf K}}
\newcommand{\vd}{{v_\Delta}}
\newcommand{\tr}{{\rm Tr}}
\newcommand{\bj}{{\bf j}}
\newcommand{\bn}{{\bf \hat{n}}}
\newcommand{\cH}{{\cal H}}
\newcommand{\cT}{{\cal T}}
\newcommand{\cM}{{\cal M}}
\newcommand{\BGamma}{{\bm \Gamma}}
\newcommand{\bsig}{{\bm \sigma}}
\newcommand{\bpi}{{\bm \pi}}

\title{Stability of Majorana Fermions in Proximity-Coupled Topological Insulator Nanowires}

\author{A. M. Cook, M. M. Vazifeh, and M. Franz}
\affiliation{Department of Physics and Astronomy,
University of British Columbia, Vancouver, BC, Canada V6T 1Z1}

\begin{abstract}
It has been shown previously that a finite-length topological
insulator nanowire, proximity-coupled to an ordinary bulk s-wave
superconductor and subject to a longitudinal applied magnetic field,
realizes a one-dimensional topological superconductor with an unpaired
Majorana fermion (MF) localized at each end of the nanowire. Here, we study the stability of these MFs with respect to various perturbations that are likely to occur in a physical realization of the proposed device.  We show
that the unpaired Majorana fermions persist in this system for any
value of the chemical potential inside the bulk band gap of order 300
meV in Bi$_2$Se$_3$ by computing the Majorana number.  From this calculation, we
also show that the unpaired Majorana fermions persist when the
magnetic flux through the nanowire cross-section deviates
significantly from half flux quantum. Lastly, we demonstrate that the
unpaired Majorana fermions persist in strongly disordered wires with
fluctuations in the on-site potential ranging in magnitude up to several times the size of the bulk band gap. These results suggest this solid-state
system should exhibit unpaired Majorana fermions under accessible conditions likely important for experimental study or future applications.
\end{abstract}
\maketitle

\section{Introduction}

In 1937, Ettore Majorana first showed that the complex Dirac equation can be separated into a pair of real wave equations, each of which is satisfied by real fermionic fields \cite{majorana}.  Such a real fermionic field, denoted by $\Psi$, satisfies the property that $\Psi=\Psi^\dagger$.  A particle created by this field, known as a Majorana fermion, is therefore distinguished by the fact that it is its own antiparticle \cite{wilczek1,franz1,alicea,majreview}.  Having many properties which make them interesting from the standpoint of fundamental science, while also being a possible
platform for fault-tolerant, scalable quantum computation \cite{a12,a42,kitaev2,a44,a45,a46}, Majorana fermions are of tremendous interest to the condensed matter community.  After intense effort, some proposals to realize Majorana zero-modes made in recent years\cite{sau2,oreg1} seem to be bearing fruit, with signatures of Majorana fermions already being reported\cite{be150,be142,furdyna1}.  Of the many devices proposed for harbouring Majorana fermions \cite{alicea,majreview}, however, virtually all face considerable experimental challenges in achieving the conditions necessary for Majorana fermion emergence. Additional hurdles are associated with the control and  manipulation of MFs which is necessary for harnessing their potential for quantum computation.  Thus, even if Majorana fermions have indeed been conclusively observed, there remains a need for more accessible platforms with which to realize MFs sufficiently robust for applications.  

\begin{figure}[t]
\includegraphics[width=7cm]{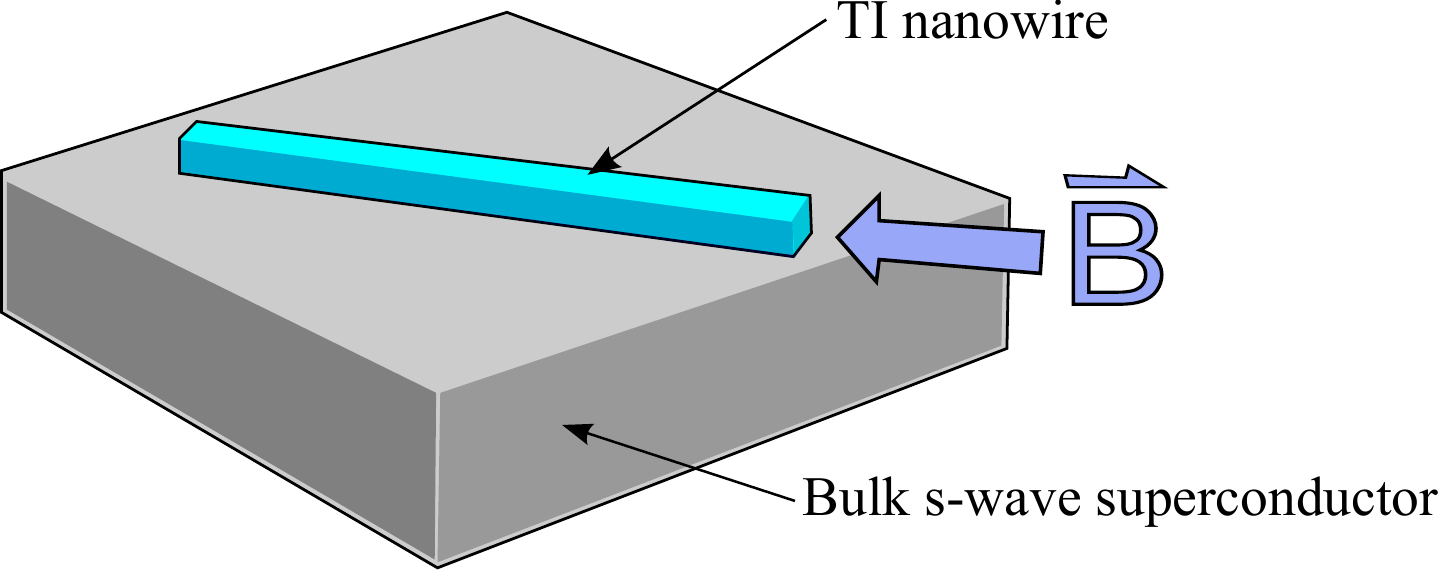}
\caption{Schematic of the proposed device. Magnetic field $\bB$ is applied along the axis of the wire taken to coincide with the $z$-direction. } \label{devicesetup}
\end{figure}

The purpose of this paper is to present results on the stability of MFs in a solid-state device previously predicted \cite{cook} to host these quasiparticle excitations.  The device, depicted schematically in Fig. \ref{devicesetup}, consists of a nanowire fashioned out of a strong topological insulator (STI), such as Bi$_2$Se$_3$ or Bi$_2$Te$_2$Se, placed in contact with an ordinary s-wave superconductor (SC), subject to an applied magnetic field along the axis of the nanowire.  We show that MFs are remarkably stable in this device, making it unique amongst the many proposals for observing MFs in solid-state systems and a significant advancement towards study of MFs and development of MF-based technology. 

We note that Bi$_2$Se$_3$ nanowires and nanoribbons have been synthesized and can exhibit diverse morphologies controllable by growth conditions\cite{kongnanoribbon}.  Aharonov-Bohm (AB) oscillations in the longitudinal magneto-resistance of Bi$_2$Se$_3$ nanoribbons have also been observed, proving the existence of a coherent surface conducting channel\cite{pengnanoribbon}.  Studies of magneto-resistance of Bi$_2$Se$_3$ nanoribbons under a variety of magnetic field orientations also reveal a linear magneto-resistance that persists to room temperature and is consistent with transport through topological surface states\cite{tangnanoribbon}.  Lastly, the superconducting proximity effect and possible evidence for Pearl vortices has been observed in Bi$_2$Se$_3$ nanoribbons\cite{zhangnanoribbon}.  This experimental progress suggests our proposed device may be realized experimentally with relative ease.

Stability of MFs in our proposed device is confirmed by showing that the degenerate quasiparticle ground state is separated from excited states by an energy gap close to the superconducting (SC) gap which can be as large as $\sim$
10 meV, through study of a low-energy analytical theory and numerical study of a lattice-model Hamiltonian.  We also compute the topological phase diagram for the system numerically to show that MFs exist in the system for any value of the chemical potential in the bulk band gap of the TI (for Bi$_2$Se$_3$, $\sim$ 300 meV).  Furthermore, we find that the topological phase corresponding to the presence of MFs persists even when the chemical potential is in the bulk conduction band, although, since we observe rapid collapse of the excitation gap in this regime, this result is of limited experimental relevance.  

These results also support additional explicit numerical studies of the robustness of MFs against non-magnetic disorder also discussed, which show that the MFs persist in the presence of fluctuations in the on-site chemical potential in an explicit lattice model Hamiltonian on the order of the bulk band gap (300 meV).  As such, previous expectations that MFs would be robust against non-magnetic disorder according to Anderson's theorem, because time reversal symmetry (TRS) holds in this device under operating conditions \cite{cook}, are here confirmed.  

In the most promising of other proposals, the MFs are protected by an SC gap of at most 1 meV, and the chemical potential must also be tuned to lie within a window of the same size\cite{sau2,oreg1}.  Although these requirements are possible to achieve in experiments on individual wires \cite{be150}, such fine tuning will be difficult to replicate in more complex setups, i.e.\ those containing wire networks necessary for MF manipulation\cite{alicea77}.  Furthermore, other proposals are not predicted to possess MFs under TR-invariant conditions, meaning these devices are not expected to be robust against non-magnetic disorder.  Therefore MFs  constructed in these proposals might be too delicate to be useful in practical applications.  Our results on the remarkable stability of MFs in the TI nanowire-based proposal therefore outline a practical route towards applications based on the physics of Majorana fermions.
\section{Majorana Fermions}
\subsection{Significance and relevance of Majorana fermions}
In the more than 70 years since Majorana's seminal paper\cite{majorana}, it is only very recently that any experimental evidence of Majorana fermions has been obtained.  Recent developments in topological states of matter have potentially already led to successful construction of Majorana fermions in solid-state systems\cite{be142,be150}, as quasiparticle excitations\cite{majreview}.  The race to experimentally study these quasiparticles has been especially heated because of the observation by Kitaev \cite{kitaev1} that Majorana fermions could be used as a platform for robust quantum computing.  In this scheme, a quantum bit is stored in a Dirac fermion that has been teased apart into two Majorana fermions.  If these two Majorana fermions are separated spatially from one another, then, whether this shared fermion is occupied or empty, it is distributed non-locally, and no local perturbation can measure this shared quantum bit \cite{kitaev1}.

Furthermore, a system of $N$ spatially-separated Majorana fermions is predicted to satisfy non-Abelian statistics, implying that such a system has an $N$-quasiparticle ground state that is degenerate.  This degeneracy allows adiabatic interchange of the quasiparticles, or braiding, to correspond to unitary operations on the ground state.  For Majorana fermions, it has also been shown that the only way to perform unitary operations on the ground state - which could be used for computing - is by braiding, and these operations are dependent only on the topology of the braid.  Since the system is in a topological phase when Majorana fermions are present, this degenerate ground state is also separated from the rest of the spectrum by an energy gap known as the "minigap".  If the temperature is much lower than the minigap, and the system is weakly perturbed using frequencies much smaller than the gap, the system evolves only within the ground state subspace \cite{nayak1}.

All of these features combined mean that a system of spatially-separated Majorana fermions could be used as a quantum computer that is immune to the tremendous obstacle faced by most other proposed platforms for quantum computing known as decoherence \cite{shor}.  Experimental confirmation of the existence of Majorana fermions is a crucial first step towards practical quantum computing, but it is imperative that platforms possessing robust Majorana fermions under stable conditions be identified and developed.
\subsection{Other existing proposals for realizing Majorana fermions experimentally}
There is no shortage of proposals for realizing Majorana fermions experimentally.  Earlier suggestions for physical systems that support Majorana fermion states include fractional quantum Hall states at filling $\nu={5\over2}$\cite{or2} and Helium-3\cite{a8}.  These ground-breaking proposals are thought to be extremely challenging to realize experimentally\cite{oreg1}, however.  We will discuss the many other proposals and comment on the experimental challenges they face below.

2D topological insulators have long been proposed as platforms for realizing Majorana zero-modes, for instance, having the advantages of greatly facilitating Josephson-based Majorana detection, long considered to be smoking gun confirmation of the presence of Majorana zero-modes\cite{alicea}, as well as being unaffected by non-magnetic disorder due to time-reversal invariance\cite{a94,a105} and, in principle, possessing a large pairing gap exhibited by the parent superconductor\cite{a89,a94}.  However, of many materials predicted to be 2D topological insulators\cite{a84,a104,a106,a107,a108,a109,a110}, only one, HgTe, has been confirmed experimentally thus far\cite{a111, a112}, although there has also been some evidence recently that InAs/GaSb quantum wells may also exhibit a topological insulator phase\cite{a113,a114}.  2D semiconductor heterostructures have also shown promise as platforms for realizing Majorana zero-modes, but face challenges due to small spin-orbit energies\cite{a206,a207}, a need for difficult-to-engineer, high-quality interfaces, and limited tunability\cite{alicea}.  

An innovative proposal for realizing Majorana zero-modes in three dimensional topological insulators due to Fu and Kane exists\cite{a49}, but this proposal, while ground-breaking, faces considerable challenges given that time-reversal symmetry must be broken to achieve Majorana zero-modes, making the device vulnerable to non-magnetic disorder\cite{alicea}.  There have also been many proposals based on Su$_2$RuO$_4$, but even in the simplest of these proposals, the minigap protecting Majorana zero-modes from excited states is in the milliKelvin range\cite{alicea}, and a beautiful proposal for realizing Kitaev's 1D toy model along an ordinary ${hc}\over{2e}$ vortex line threading a layered spinful $p+ip$ superconductor likely to be Su$_2$RuO$_4$ currently faces the same problem\cite{a178,alicea}.

There is great interest in realizing Majorana zero-modes in one dimensional systems, because they have generally been predicted to remain separated from excited states by a larger energy gap than in other proposals\cite{alicea}.  Conventional 1D wires with sizeable spin-orbit coupling, proximate to s-wave superconductors, and subject to modest magnetic fields\cite{sau2,oreg1} are seen as very promising platforms for first experimental realization of Majorana zero-modes\cite{alicea}.  These proposals must overcome numerous issues, however, such as positioning of the chemical potential in a rather small interval of roughly $1$ K over distances long compared to the wire's coherence length\cite{alicea}.  This constraint could be relaxed by applying larger magnetic fields, but this introduces other difficulties\cite{alicea}.  Tuning of the chemical potential could likely be even more difficult due to disorder-induced fluctuations in the chemical potential, since the topological phase corresponding to the presence of Majorana zero-modes appears only at finite magnetic fields in these devices, so Anderson's theorem does not protect the gap against non-magnetic disorder, which is always pair-breaking according to many previous studies\cite{a94,a97,a119, a120, a121, a122, a123, a124, a125}.  Further, since the ratio of Zeeman energy to spin-orbit energy is small for both wires made of InAs and InSb\cite{a131}, disorder is likely to play a non-trivial role\cite{alicea}.  Although there have been efforts to ameliorate this issue by eliminating an applied magnetic field\cite{a99,a133} from the device or reducing it\cite{a134,a103}, these approaches can also lead to complications that can potentially cause the Majorana zero-modes to disappear\cite{a99,a134}.

The above conventional 1D wire proposals further face the challenge that multiple sub-bands are usually occupied in these wires and gating into the lowest sub-band regime is potentially non-trivial, especially if these wires are in close proximity to a superconductor as proposed\cite{alicea}.  Multichannel wires have been shown to support the 1D topological superconductor state leading to Majorana zero-modes away from the lowest sub-band limit\cite{a121,a124,a135,a143}, but these systems still require some degree of gating, leading to proposals of increasing complexity involving regular arrays of superconducting islands in contact with the wire\cite{a146, a147, a148}.  Such work has even led to the ingenious proposal of a chain of quantum dots that would be bridged by superconducting islands\cite{a148}, but reaching the regime where only a comparatively small number of quantum dots would be needed would require very fine-tuning that would likely suffer from strong randomness\cite{alicea}.  Carbon nanotubes, also suggested as hosts for Majorana zero-modes, face considerable challenges in reaching the spinless regime with proximity-induced pairing required\cite{a151, a152, a153}, while proposals involving half-metallic ferromagnetic wire also face challenges, such as the need to couple to non-centrosymmetric superconductors with spin-orbit coupling\cite{a154, a155}. 

Despite these challenges, there have been promising experimental results for Majorana fermions based on some of these proposals.  Josephson effects at the surface of a variety of 3D topological insulators with superconducting electrodes have been observed \cite{be138,be139,be140,be141,be142,be143, be144}. While these experiments, and related Andreev conductance measurements\cite{be145,be146,be147,be148} show interesting and unusual features, these cannot be readily attributed to the single Majorana zero-mode (typically only one out of 10$^5$ modes)\cite{majreview}. 

The nanowire-based proposal of Lutchyn et al. \cite{sau2,oreg1} and Oreg et al. \cite{oreg1} has also led to convincing evidence for a Majorana zero-mode in an InSb nanowire as reported by Kouwenhoven and his group \cite{be150}.  Since then, theoretical work\cite{plee_sept} by Patrick Lee and collaborators has indicated that, under conditions for semiconducting wires with modest amounts of disorder relevant to Kouwenhoven's work, Majorana end-states are destroyed and do not give rise to quantized zero-bias peaks (ZBPs). At finite temperatures, furthermore, ZBPs of a non-topological origin are predicted to appear, leading to clusters of low-energy states localized near the wire end. These non-topological ZBPs are further anticipated to be typically stable with respect to variations of chemical potential and magnetic field, and appear and disappear under nearly identical conditions to those of true Majorana peaks. This work suggests caution is required in interpreting recent experiments to observe MZMs and that substantially longer and cleaner wires are required to conclusively observe MZMs.  

However, work by Tewari and Stanescu\cite{tewari_sept} also indicates, for a smooth confinement potential at the ends of a semiconductor Majorana wire, emergence of zero bias conductance peaks corresponding to the topologically trivial phase is necessarily accompanied by a signature similar to closing of the bulk band gap.  This gap closing signature in the end-of-wire local density of states was absent in the Kouwenhoven study, suggesting Kouwenhoven's group and others\cite{ig1,ig2} may have been successful in observing MZMs. If indeed Majorana zero-modes have finally be observed, however, there still remains a need for devices in which MFs can be realized under more accessible conditions, are robust, and can finally be manipulated for topologically-protected computation, motivating the results we present here on a proposal in which Majorana fermions occur under a wide-range of accessible conditions robustly.

\section{TI nanowire with magnetic and superconducting order}
\subsection{Low-energy theory: normal state}
We begin by presenting the low-energy analytical theory of the device \cite{cook} in greater detail to facilitate later discussion of the novel results on stability, as this foundation is later used to understand the new results.

First, we motivate the proposal with study of a cylindrical TI nanowire proximity-coupled to a bulk s-wave SC as the greater symmetry of this system permits analytical study of the low-energy fermionic excitations on the surface of the nanowire.

The low-energy fermionic excitations on the surface of the topological insulator are governed by the Dirac
Hamiltonian \cite{mirlin1}
\begin{equation}
h_0={{v}\over 2}\bigr[\hbar\nabla\cdot\bn +\bn\cdot(\bp\times\bs) + (\bp\times\bs)\cdot\bn\bigl],\label{h01}
\end{equation}
where $\bn$ is a unit vector normal to the surface, $\bp=-i\nabla$ is the momentum operator and $\bs$ is the vector of Pauli matrices in the spin space. 
We will also include the effect of a magnetic coating on the TI nanowire by adding an additional term, $h_{m} = \bs\cdot\bmm$, to the Hamiltonian.  Later, we will show that this term is not necessary for Majorana zero-modes to emerge in the device, but its inclusion will be convenient in calculations.
\begin{figure}[b]
\includegraphics[width=1\columnwidth]{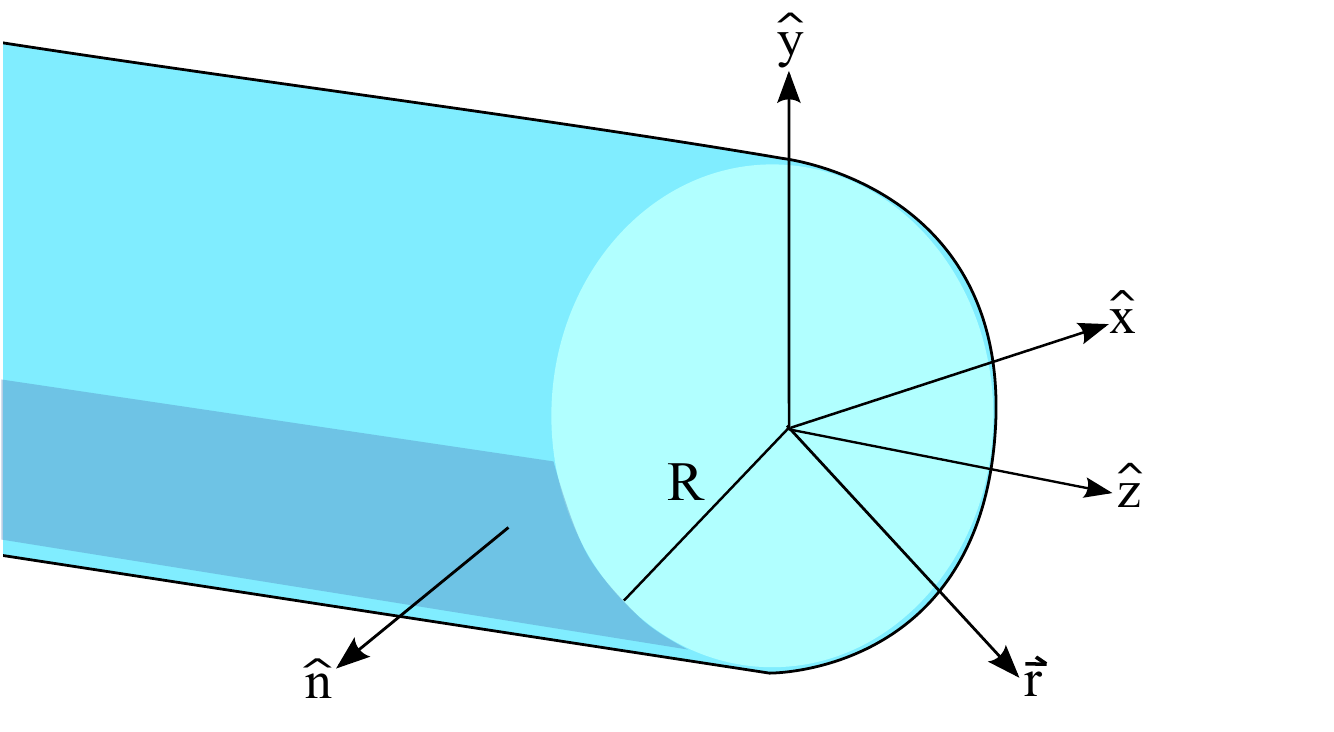}
\caption{Schematic of the device simplified for analytical study, in which a cylindrical TI nanowire is substituted for a more realistic TI nanowire with square cross-section. Magnetic field $\bB$ is still applied along the axis of the wire taken to coincide with the $z$-direction. } \label{fig1_wire_axes}
\end{figure}

Let us now consider the specific case of a cylindrical topological insulator nanowire of radius $R$ with magnetic field ${\bf B}$ applied along the $\hat{z}$-axis
as shown in Fig. \ref{fig1_wire_axes}.  The unit vector ${\bn}$ is then taken to be normal to the curved surface of the 
nanowire.  To include a flux $\Phi$ through the end of the wire (in the $\hat{z}$ direction) as proposed, 
we replace the momentum operator $\bp$ in Eq.\ (\ref{h01}) with $\bpi=\bp-(e/c)\bA$, where
$\bA=\eta\Phi_0(\hat{z}\times\br)/2\pi r^2$ is the vector potential and $\Phi_0$ is the flux quantum.  Therefore, suppressing $v\hbar$, we now have
the Hamiltonian,
\begin{equation}
h={1\over {2r}}\mathbb{I}+\bigr(\bn\times\bpi\bigl)\cdot\bs+\bs\cdot\bmm.\label{h3}
\end{equation}
Taking $\bmm = m\hat{z}$, we can rewrite the Hamiltonian in cylindrical coordinates as
\begin{equation}
h={1 \over 2R}\mathbb{I}+s_1k\sin(\phi)-s_2k\cos(\phi)-s_3\left({i \over R}\partial_{\phi}+{\eta \over R}\right)+ms_3.\label{hcyl4}
\end{equation}
To diagonalize this Hamiltonian for an infinitely long wire, we exploit the translational and rotational symmetries and write a solution $\psi_{kl}$ of the form
\begin{equation}\label{psi1} \psi_{k}(z,\varphi)=e^{i\varphi l}e^{-ikz}\begin{pmatrix} f_{kl} \\
e^{i\varphi}g_{kl} \end{pmatrix} 
\end{equation}
With this ansatz, our Hamiltonian is
\begin{equation}\label{dir2}
\tilde h_{kl}=s_2k+s_3[(l+{1\over2}-\eta)/R+m]. 
\end{equation}
The spectrum $E_{kl}$ for $m=0$, if $v\hbar$ is reinstated, is then
\begin{equation}\label{ekl} 
E_{kl}=\pm v\hbar\sqrt{k^2+{(l+{
1\over2}-\eta)^2\over R^2}}.
\end{equation}
Here $k$ labels momentum eigenstates along the cylinder while $l=0,\pm1,\dots$ is the angular momentum.
We see that the spectrum has a gapless branch for $\eta = n+{1 \over 2}$, where $n$ is any integer ($\eta=\Phi/\Phi_0$ measures the magnetic flux through the wire cross section in the units of flux quantum $\Phi_0=hc/e$).  The
periodicity $\eta\to \eta+n$ with $n$ integer in Eq.\ (\ref{ekl}) reflects the expected $\Phi_0$-periodicity in the total flux.
\begin{figure}[t]
\includegraphics[width=1\columnwidth]{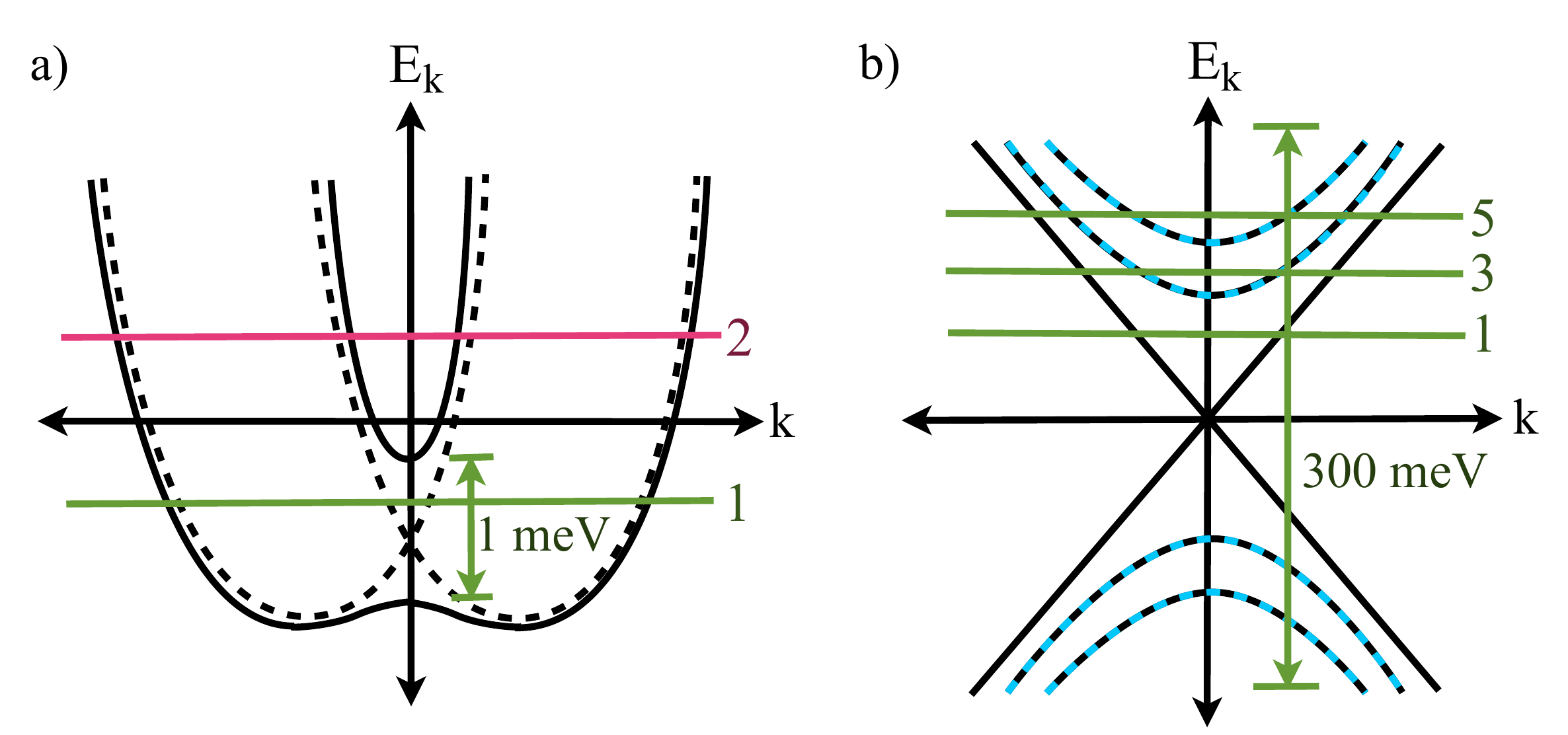}
\caption{The normal state dispersion under conditions required for emergence of Majorana fermions for (a) the Rashba-coupled semiconductor quantum wire proposal in \cite{sau2,oreg1}, where the dispersion is shown without Zeeman coupling (dashed lines) and with Zeeman coupling, and (b) our topological insulator nanowire proposal, with doubly-degenerate bands shown as black and blue dashed lines.  Green and pink horizontal lines represent the level of the chemical potential and a number to the right of a line indicates the number of Fermi points in the right half of the Brillouin zone at that value of the chemical potential.  Vertical green lines indicate the interval inside which the chemical potential can be tuned to yield Majorana fermions in the corresponding SC state.} \label{ekvsk_variouseta}
\end{figure}

We now notice that, although the branches of $E_{kl}$ are doubly-degenerate for $\eta=0$, the degeneracy is lifted for $\eta\neq0$.
One can always find a value of the chemical potential $\mu$ that yields a single pair of non-degenerate Fermi points for $\eta\neq0$, 
as illustrated in Fig. \ref{ekvsk_variouseta}, or more generally, an odd number of such pairs. According to the Kitaev's criterion \cite{kitaev1} pairing induced by the proximity effect is then expected to drive the system into a topological phase. In the special case $\eta=1/2$ the two lowest bands are non-degenerate, while all higher bands are doubly-degenerate, yielding an odd number of Fermi points at any $\mu$ in the bulk band gap. 
  While the semiconductor wire proposal \cite{sau2,oreg1} only possesses Majorana fermions for values of the chemical potential in a ~1 meV interval, our proposal possesses Majorana fermions for any value of the chemical potential $\mu$ inside the 300 meV bulk band gap of Bi$_2$Se$_3$.  Fine-tuning of the chemical potential is unnecessary in our device at $\eta=1/2$ due to the specific pattern of degeneracies of the bands which is in turn protected by the Kramers theorem.

The surface Dirac Hamiltonian (\ref{h01}) is expected to be valid in
the limit when the surface state penetration depth $\zeta$ is much
smaller than the wire radius $R$. In the opposite limit of a thin
wire, $\zeta> R$, one could worry that the wavefuncion overlap in the
interior of the wire might lead to the formation of a gap, as happens
e.g.\ in thin TI films. We study the limit of a thin wire in Appendix
A, based on a 3D effective model of a TI. The results of this study
are interesting. We find that the gapless mode actually persists for an
arbitrary radius $R$ in the case when the magnetic flux $\Phi_0/2$ has
the form of a $\delta$-function centered at the axis of the
cylinder. For a uniform magnetic flux a gap in the surface state opens
up and its magnitude is proportional to $(\zeta/R)^2$ representing the
amount of magnetic flux to which the wave function is exposed. Thus,
in the case of a cylindrical wire, it is not the wavefunction overlap
(which would lead to a gap $\sim e^{-R/\zeta}$) but the amount of
$\cT$-breaking in the system that determines the gap. We note that our
numerical simulations discussed in Sec.\ IV below indicate that for
moderately thin wires, i.e.\ $\zeta$ being a significant fraction of
$R$, the gapless state actually persists but now exists at somewhat
higher magnetic flux. Also, the wires likely to be used in an experiment\cite{kongnanoribbon,pengnanoribbon,tangnanoribbon}
are tens of nm thick and are thus in the thick-wire limit, $\zeta$ being typically just a few lattice spacings.

Finally we note that the surface Dirac Hamiltonian (\ref{h01})
represents the simplest possible model that neglects anisotropies
present in real materials, such as Bi$_2$Se$_3$. Such anisotropies
will necessarily lead to a spin texture that depends on the
crystallographic direction of the surface which can be described by an
effective $4\times 4$ surface Hamiltonian.\cite{mele1} It would be
interesting to understand these effects for various wire geometries
but we leave this problem for future study. We note that our lattice
Hamiltonian employed in Sec.\ IV includes the above mentioned
anisotropies and the results based on it confirm all the essential
features of our simple analytical model.

\subsection{Low-energy theory: Majorana fermions}

To study the emergence of Majorana fermions in the simplest possible setting, we now focus on the $\eta={1 \over 2}$ case and consider values
of the chemical potential satisfying $|\mu| < {v\hbar \over R}$, i.e. intersecting only the $l=0$ branch of the spectrum.  The
Hamiltonian for this branch then becomes $h_k = ks_2-\mu+ms_3$, where we have explicitly included the chemical potential term.
The Bogoliubov-de Gennes Hamiltonian describing the proximity-induced superconducting order in the nanowire can be written, in the second-quantized notation, as 
$H=\sum_k\Psi^\dagger_k\cH_k\Psi_k$ with
$\Psi_k=(f_k,g_k,f^\dagger_{-k},g^\dagger_{-k})^T$ and
\begin{equation}\label{hsc1}
\cH_k=\begin{pmatrix} h_k & \Delta_k \\
-\Delta^*_{-k} & -h^*_{-k}
 \end{pmatrix}.
\end{equation}
For the surface state, $\eta=1/2$ represents a ${\cal
T}$-invariant point at which $h^*_{-k}=h_k$.  Therefore, $\cH_k$ can be written as
\begin{equation}\label{hsc2}
\cH_k=\begin{pmatrix} h_k & \Delta_k \\
-\Delta^*_{-k} & -h_k
 \end{pmatrix}.
\end{equation}
 In the following, we consider the simplest s-wave pairing potential $\Delta_k=\Delta_0i\bs_2$, with $\Delta_0$ a (complex) constant order parameter,
which corresponds to the pairing term $\Delta_0(f^\dagger_{k}g^\dagger_{-k}-g^\dagger_{k}f^\dagger_{-k})$. 
It is useful to note that this form of $\Delta_k$ actually implies a vortex in the SC order parameter, as can be seen by transforming $\cH_k$ back into the original electron basis, i.e.\ undoing the transformation indicated in Eq.\ (\ref{psi1}). The phase of the order parameter in this basis winds by $2\pi$ on going around the cylinder as required in the presence of the applied magnetic field whose total flux is $\Phi_0/2$.
  
Introducing
Pauli matrices $\tau_\alpha$ in the Nambu space and assuming
$\Delta_0$ real, we can write
\begin{equation}\label{hsc3}
	\cH_k = \tau_3(ks_2-\mu)+\tau_3s_3m-\tau_2s_2\Delta_0.
\end{equation}
(Here we have again taken $v=\hbar=1$.)  The spectrum for this Hamiltonian is 
$E_k = \pm(k^2+\mu^2+m^2+\Delta_0^2\pm2(k^2\mu^2+\mu^2m^2+m^2\Delta_0^2)^{1/2})^{1/2}$.  We now consider a special case when $\mu=0$. The Hamiltonian simplifies, $\cH_k=\tau_3s_2k+\tau_3s_3m-\tau_2s_2\Delta_0$ and the spectrum assumes a simple and suggestive form
\begin{equation}\label{Esc2}
	E_k=\pm\sqrt{k^2+(m\pm\Delta_0)^2}.
\end{equation}
We observe that the spectrum is fully gapped in the presence of either SC or magnetic order but has a gapless branch when $m=\pm\Delta_0$.  Thus, we expect a topological phase transition at this point. Consequently, we expect gapless modes to exist at an interface between SC and magnetic domains in a wire.
\begin{figure}[t]
\includegraphics[width=1\columnwidth]{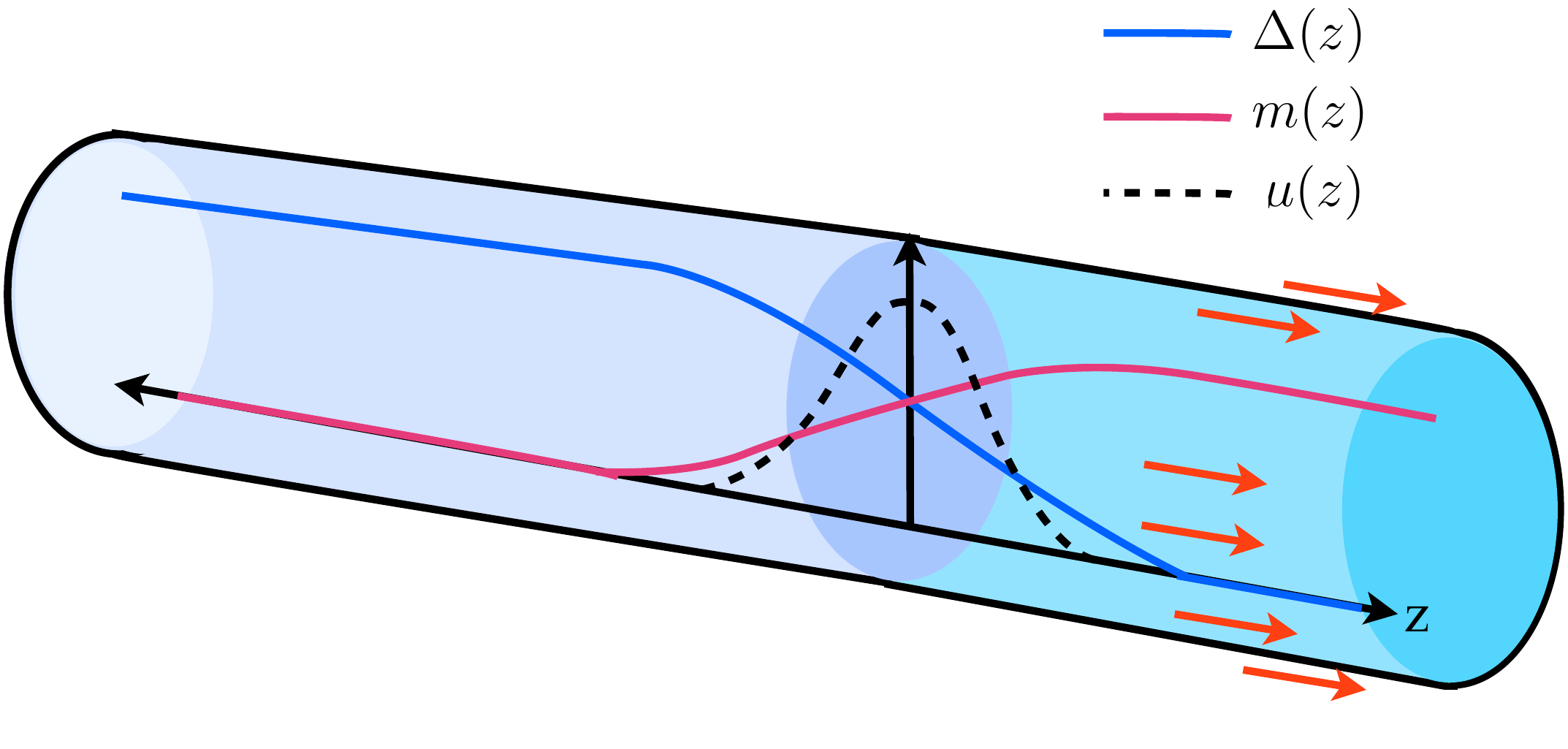}
\caption{A convenient possible choice for the SC/magnetic domain wall at $z=0$.  $\Delta(z)$ is the SC order parameter and $m(z)$ the magnetic order parameter.  A dashed line shows the zero-mode solution $\mu(z)$ for this domain wall.  This particular choice of boundary conditions can be used to show Majorana fermions occur at the ends of the TI nanowire irrespective of precise boundary conditions.} \label{fig3spatialvarordparams}
\end{figure}

Now consider spatially varying $m(z)$ and $\Delta(z)$ such that $m(0)=\Delta(0)$ as sketched in Fig. \ref{fig3spatialvarordparams}.  With these choices for the order parameters, we expect the spectrum to be gapped far away from the domain wall, but we expect gapless modes localized near $z=0$.  To determine if there are any fermionic zero modes, we rotate $\cH_k$ in $s-\tau$ space so that the rotated Hamiltonian is completely off-diagonal.  That is, we work with $\tilde{\cH}_k=U\cH_kU^{-1}$, where 
\begin{equation}\label{U1}
U=e^{-i{\pi \over 4}s_2}e^{-i{\pi\over4}\tau_2}.
\end{equation}
Then $\tilde{\cH}_k=\tau_1s_2k+\tau_1s_1m-\tau_2s_2\Delta$, so $\tilde{\cH}_k$ is of the form
 \begin{equation}\label{hsctilde1}
\tilde{\cH}_k=\begin{pmatrix} 0 & D_k \\
D^\dagger_k & 0
 \end{pmatrix},
\end{equation}
where $D_k=s_2k+s_1m+is_2\Delta$.

We now replace $k\to -i\partial_z$ and look for solutions $\tPsi(z)$ satisfying
\begin{equation}\label{HPsiequalszero}
\tilde{\cH}_k\tPsi(z)=0.
\end{equation}
Taking $\tilde{\Psi}(z)=\left(\psi_1,\psi_2,\psi_3,\psi_4\right)^T$ and reinstating $v$, Eq.\ (\ref{HPsiequalszero}) yields four independent equations:
\begin{align}
    (+v\partial_z+m+\Delta)\psi_1&=0\label{psi1eqn} \\
    (-v\partial_z+m-\Delta)\psi_2&=0\label{psi2eqn} \\
     (+v\partial_z+m-\Delta)\psi_3&=0\label{psi3eqn} \\
     (-v\partial_z+m+\Delta)\psi_4&=0\label{psi4eqn}
\end{align}
Here we have suppressed the $z$ dependence.  The solution $u(z)$ of an equation of the form
\begin{equation}\label{jrzeromode1}
\left[v\partial_z+\omega(z)\right]u_z=0,
\end{equation}
a Jackiw-Rossi zero mode \cite{jackiw1}, can be written as
\begin{equation}\label{jrzeromode2}
u(z)=u_0e^{{-1\over v}\int_0^zdz'\omega(z')}.
\end{equation}
This solution is normalizable provided that $\omega(z)$ has a soliton profile, i.e.\ is proportional to ${\rm sgn}(z)$ for large $|z|$.
According to our assumptions, $m(z)+\Delta(z)>0$ for all values of $z$, so there is no normalizable solution for $\psi_1$ or $\psi_4$.  With $v>0$, there is also no normalizable solution for $\psi_2$, but there is one for $\psi_3$:
\begin{equation}\label{psi3soln}
\psi_3(z)=u_0e^{{-1\over v}\int_0^zdz'\left(m(z')-\Delta(z')\right)}.
\end{equation}
(For $v<0$, $\psi_2$ would be the normalizable solution instead.)  Thus, for $\Delta(z)$, $m(z)$ as given in Fig. \ref{fig3spatialvarordparams}, our Hamiltonian has a single zero-mode solution of the form $\tilde{\Psi}_0=(0, 0, 1, 0)^Tu(z)$ localized near the domain wall at $z=0$.  This solution is valid as long as $v>0$ and $m(z)-\Delta(z)\rightarrow\pm$const for $z\rightarrow\pm\infty$.
To see if the zero mode $\tPsi_0(z)$ corresponds to a Majorana fermion, we undo the unitary rotation and inspect the corresponding solution $\Psi_0(z)=U^{-1}\tPsi_0(z)$, which is
\begin{equation}\label{soln1}
{\Psi}_0(z)={1\over2}(1, -1, 1, -1)^Tu(z).
\end{equation}
In second quantization, the field operator destroying the particle in the state $\Psi_0(z)$ is
\begin{equation}\label{soln2}
\hat{\psi}_0={1\over2}\int dzu(z)\left[f(z)-g(z)+f^\dagger(z)-g^\dagger(z)\right].
\end{equation}
where $f(z)$, $g(z)$ are real-space versions of the $f_k$,  $g_k$ operators in $\Psi_k$ .  Since $u(z)$ is real, it holds that $\hat{\psi}_0^\dagger=\hat{\psi_0}$, so $\Psi_0$ represents a Majorana fermion.

With a few additional observations, the above calculation can be used to show that an additional unpaired Majorana mode exists at the SC end of the wire irrespective of boundary conditions.  First, recall that in a finite system Majoranas always come in pairs, since they are formed from ordinary fermions \cite{majreview}.  This second Majorana fermion, being a zero-mode, cannot not exist in the nanowire bulk where the spectrum is gapped. It cannot exist at  the magnetic end because the magnetic order does not support the requisite particle-hole mixing. The second MF must therefore be at the SC end of the nanowire, irrespective of the exact boundary condition.  From this, we can argue that the special conditions used to establish the existence of unpaired MFs in the device are unnecessary:  The zero-modes in fact exist in the device under generic boundary conditions as confirmed by explicit numerical study using a lattice model, discussed in section III.  A specific example of Majorana end-states obtained in such a lattice calculation under general conditions is given in Sec. IV.A below.

\subsection{Energy gap protecting Majorana zero-modes}
As mentioned in \cite{returnminigap}, in order to detect and manipulate MFs under experimentally accessible conditions it is crucial that they are protected from all other excitations by a gap. The latter is often refered to as a `minigap' because typically there will be other excitations inside the bulk gap. We study the minigap in this TI nanowire-based device both analytically and numerically.  

In this section we estimate the minigap for the superconducting TI nanowire using the analytical low-energy theory.  Specifically, we wish to find the lowest non-zero eigenvalue of $\tilde{\cH}_k$ defined in Eq.\ (\ref{hsctilde1}).  We start by squaring the Hamiltonian.  We find, with $k\rightarrow-i\partial_z$ and $D_k^\dagger D_k\rightarrow D^\dagger D$,
\begin{equation}\label{DDdag} 
D^\dagger D = \partial_z^2+\left[\Delta'(z)-s_3m'(z)\right]+\left[\Delta(z)-s_3m(z)\right]^2.
\end{equation}
The two independent equations for $s_3=\pm 1$ in $D^\dagger D$ can more conveniently be written as $(D^\dagger D)_+$ and $(D^\dagger D)_-$, where
\begin{equation}\label{DDdagpm} 
\left(D^\dagger D\right)_\pm = \partial_z^2+\left[\Delta'(z)\mp m'(z)\right]+\left[\Delta(z)\mp m(z)\right]^2.
\end{equation}
To find the energy of the first excited state, we look for solutions $\psi$ satisfying
\begin{equation}\label{hamexcited1} 
h\psi=\epsilon\psi,
\end{equation}
where $h=\tilde{\cH}^2$ and  $\epsilon>0$.  We consider $m(z)$, $\Delta(z)$ such that $\Delta(z)+m(z)=$const for each $z$, and $\Delta(z)-m(z)=f(z)$ having a soliton profile, e.g. we may take $f(z)=\Delta_0\tanh \left({z/ \xi}\right)$, as shown in Fig. \ref{fig3spatialvarordparams}.  Then $(D^\dagger D)_-$ yields no bound states.  $(D^\dagger D)_+$, however, has the form, with velocity $v$ restored,
\begin{equation}\label{hamexcited2} 
(D^\dagger D)_+=-v^2\partial_z^2+vf'(z)+f^2(z)
\end{equation}
For bound-state energies much less than $\Delta_0$, $f(z)$ can be approximated as linear in the vicinity of $z=0$.  With $f(z)\simeq -\Delta_0{z\over \xi}$, where $\xi$ is the length scale over which the SC order parameter varies near the domain wall, we then have
\begin{equation}\label{hamexcited3} 
(D^\dagger D)_+=-v^2\partial_z^2-{{v\Delta_0}\over\xi}+\left(\Delta_0 \over \xi\right)^2z^2.
\end{equation}
This is the Hamiltonian for the harmonic oscillator with the identification ${{\hbar^2}\over {2m}}=v^2$, ${{m\omega^2}\over 2}=\left({\Delta_0 \over \xi}\right)^2$, and $\hbar^2\omega^2=4v^2{\Delta_0^2 \over \xi^2}$.  Therefore, allowed eigenenergies of $(D^\dagger D)_+$ bounded above by $\Delta_0^2$ are
\begin{equation}\label{hamexcited4} 
\epsilon_n=\hbar\omega(n+{1\over2})-{{v\Delta_0}\over \xi}={{2\hbar v \Delta_0}\over\xi}n,
\end{equation}
where $n$ is any non-negative integer.  The energy spectrum of $\tilde{\cH}$ in this approximation is then
\begin{equation}\label{hamexcited5} 
E_n=\pm\sqrt{\epsilon_n}=\pm\sqrt{{{2\hbar v\Delta_0}\over\xi} n}.
\end{equation}
Since $\xi$ is the length scale over which the SC order parameter varies near the wire end, it is at most the SC coherence length ${{\hbar v}\over {\pi \Delta_0}}$.  The minimum energy of the first excited state $E_1$ is then
\begin{equation}\label{hamexcited5} 
E_1=\Delta_0\sqrt{2\pi},
\end{equation}
which is already greater than $\Delta_0$.  Therefore, there are no excited states where the linear approximation holds.  There can be some at energies close to $\Delta_0$ and this is consistent with numerical results presented in \cite{cook}. We thus conclude that the minigap amplitude is close to $\Delta_0$ in this case.

We also note that the calculation presented above is valid in the special case $\mu=0$. For non-zero chemical potential the situation is more complicated and we are not able to find a simple analytic solution for the excited states in this case. Since the density of states of the underlying Dirac semimetal grows with increasing energy we expect there to be more low-lying excited states when $\mu\neq 0$ and thus reduced minigap. This expectation is indeed confirmed by our numerical simulations discussed below.

\subsection{Majorana state in a finite-wire configuration}

So far in our analytical calculation we have shown that a Majorana state exists in a TI wire at the interface between SC and magnetic domains. A more realistic situation from an experimental point of view is
to consider a finite-length wire located on top of a s-wave superconductor. One would expect 
the Majorana zero modes to live close to the ends of the wire where the SC order parameter vanishes and the magnetic flux 
pierces of the surface of the wire. Our numerical results below confirm this intuition but here we briefly show that one can also analytically prove 
this for a simple wire configuration and find the Majorana state. 

\begin{figure} 
\includegraphics[width=6cm]{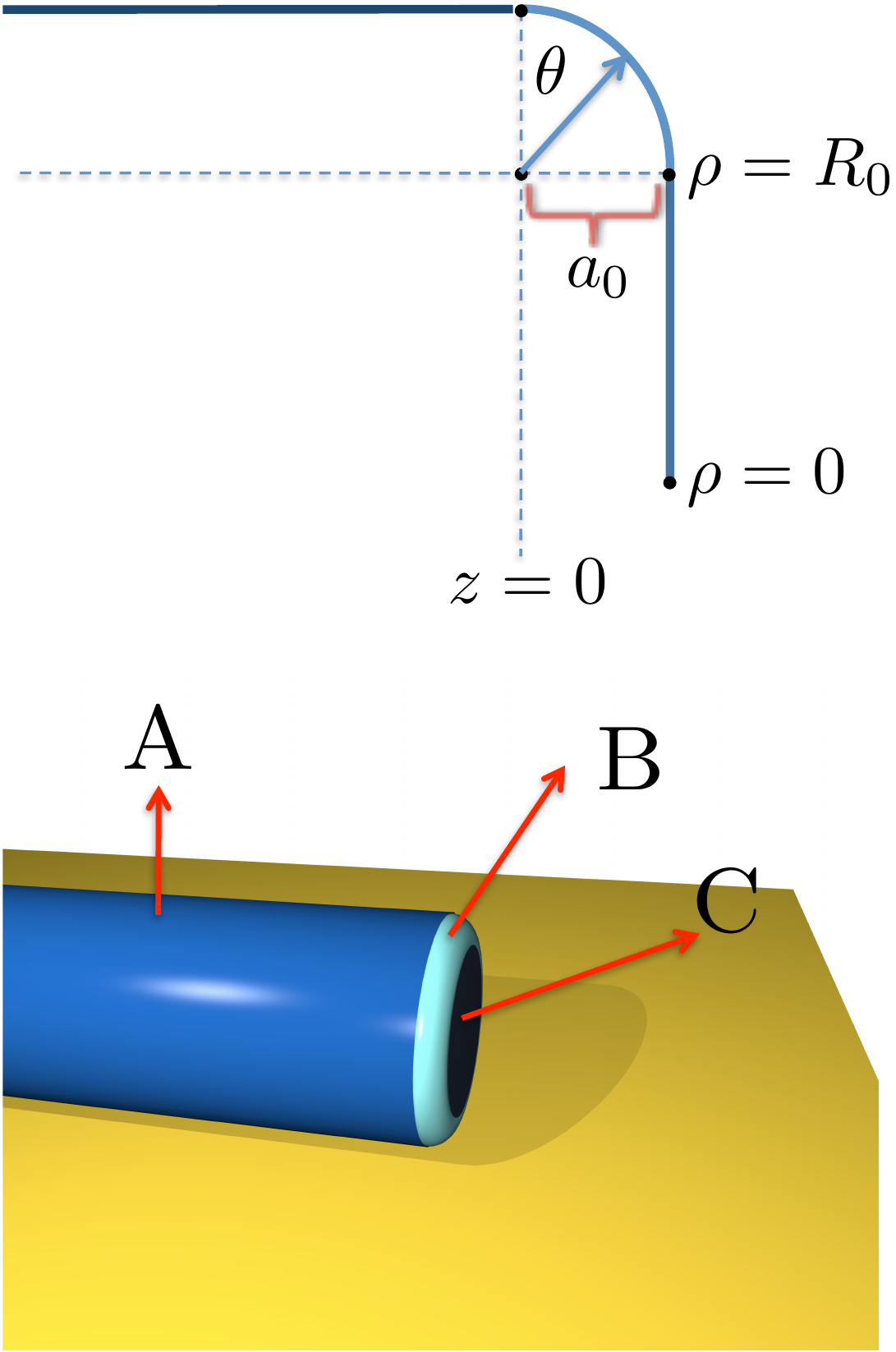} 
\caption{The schematics of a finite-length wire on a superconducting substrate. The surface of the wire has been divided into three regions $A$, $B$ and $C$ (bottom panel) according to the form of the Hamiltonian. The top panel details the assumed shape of the wire end with various quantities used in the text indicated.}\label {finite-wire3D}
\end{figure}

Consider one end of a finite wire with a configuration as shown in Fig. (\ref{finite-wire3D}): a cylinder with the sharp edge smoothed out.  
For convenience we think of the resulting surface 
separated into three regions. As before we consider a uniform 
magnetic flux with $\eta = 0.5$ through the wire but we assume that it only pierces region $C$ of the surface. The form of the Hamiltonian is different in each region as one would expect due to 
the curvature effects and is discussed in detail in  the Appendix B. The regions have been chosen in such a way that the first spatial derivative operator that appears in the Dirac Hamiltonian is 
continuous everywhere on the wire including the boundaries between the regions. This way it is legitimate to use Dirac hamiltonian Eq.(\ref{h01}) to study the surface states 
since the normal unit vector to the surface of the TI is well-defined everywhere on the wire. Note that this would not be the case if we considered sharp edges.  We can exploit the azimuthal symmetry of this 
configuration and perform the unitary transformation defined by Eq.\ (\ref{psi1}). The full BdG Hamiltonian for the $l=0$ branch then reads  
\begin{equation}
H = \int d \zeta \Psi^{\dag}(\zeta) \mathcal{H}(\zeta) \Psi^{ }(\zeta)
\end{equation}
where $\mathcal{H}(\zeta)$ is a $4\times 4$ matrix given by ($\hbar = v = 1$)
\begin{equation}
\mathcal{H}(\zeta)=\tau_3 \left[ i s_2 \partial_{\zeta} + g(\zeta) s_1 + m(\zeta) s_3 + \lambda(\zeta)\right] - \Delta(\zeta) \tau_2 s_2   \nonumber
\end{equation}
The Hamiltonian is a function of the length $\zeta$ which parametrizes the geodesic curves that connect two end points on each section of the surface. We choose $\zeta$ to be zero at the end of the cylinder (i.e.\ at the boundary between regions $A$ and $B$ as shown in Fig. (\ref{finite-wire3D})). Note that on the cylinder it is equivalent to the $z$ variable we used before. In general we have
\begin{equation}
\zeta \equiv \left \{ \begin{matrix}  z \;\;(z\leqslant 0)  \;\;\;\;\;\;\;\;\;\;\;\; \;\;\;\;\;\;\;\;\;\;\;\; \;\;\;\;\;\;\;\;\;\;\;\;  \zeta \in A \\  \\ a_0 \theta \;\;(0\leqslant \theta \leqslant \pi/2) \;\;\;\;\;\;\;\;\;\; \;\;\;\;\;\;\;\;\;\;\;\;  \zeta \in B \\ \\ {R_0 + a_0 \pi/2 - \rho } \;\;(0 \leqslant \rho \leqslant R_0)   \;\;\;\;\;  \zeta \in C     \end{matrix} \right . 
\end{equation}
where $a_0$, $R_0$ are the radii of the connecting torus and the cylinder ($a_0 \ll R_0$) respectively.
$g(\zeta)$ and $\lambda(\zeta)$ are two functions that arise due to the curvature effects. They are defined as
\begin{equation}
g(\zeta) = - \frac{1}{2 \rho} \times \left \{ \begin{matrix}  0 \;\;\;\;\;\;\;\;\;\;\;\;\;\;\;\;\;\;\;\;   &&  \zeta \in A, B \\  \\  1 - (\rho/R_0)^2  && \zeta \in C      \end{matrix} \right . 
\end{equation}
and
\begin{equation}
\lambda(\zeta) =  \frac{1}{2 a_0} \times \left \{ \begin{matrix}  0   &&    \zeta \in A, C \\  \\  1  &&   \zeta \in B      \end{matrix} \right . 
\end{equation}
Assuming that the magnetic flux is narrower than the wire cylindrical shaft we can neglect the Zeeman field in all regions except region $C$. Therefore we consider the following profile for it
\begin{equation}
m(\zeta) =  \left \{ \begin{matrix}  0   &&     \zeta \in A, B \\  \\  m_0  && \zeta \in C      \end{matrix} \right . 
\end{equation}
As mentioned previously, we expect MF to exist near the end of the wire irrespective of the details of the boundary condition. For simplicity, therefore, $\Delta(\zeta)$ is assumed to be nonzero and uniform only in the range of $\zeta$ parametrizing the cylindrical shaft of the wire and the rounded edge ($\zeta \in A, B$).  
In region $C$ we assume $\Delta(\zeta)=0$ in accord with the intuition that the SC order will be suppressed here due to the magnetic field piercing the surface and the presence of the vortex.
Finally, we consider a very long wire so that we can assume that the overlap between localized states at the two ends is negligible. This way we can look for solutions with zero energy and safely exclude the solutions that grow exponentially towards the other end. Thus, we seek the solutions of the following equation
\begin{equation}\label{zero}
\mathcal{H}(\zeta) \Psi(\zeta) = 0
\end{equation}
and investigate whether there is a solution (localized near the end) which satisfies the Majorana condition $\hat{\Psi}_0^{\dag} = \hat{\Psi}_0$. 

\begin{figure} 
\includegraphics[width=8cm]{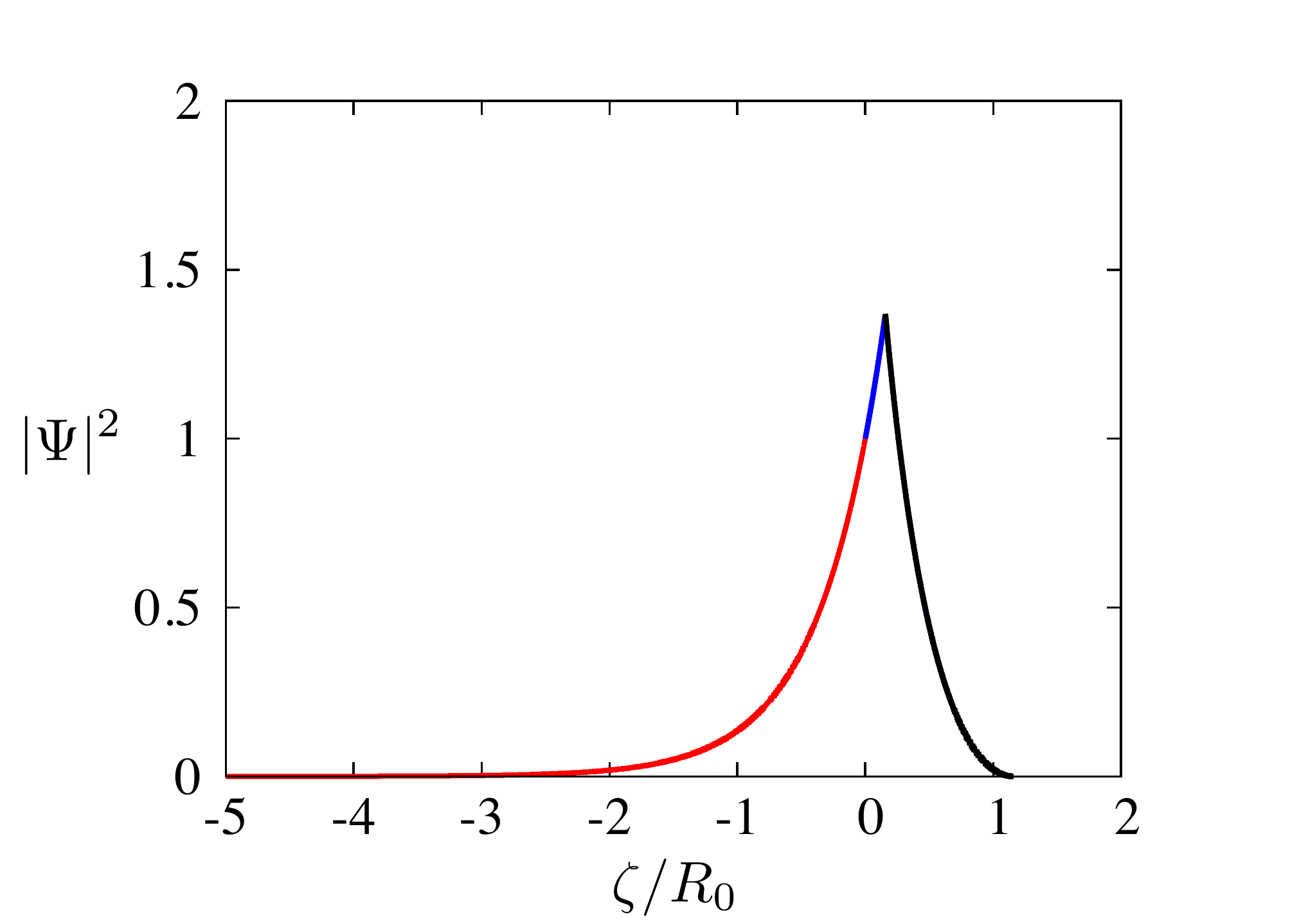} 
\caption{ The probability density of the Majorana state (up to a normalization constant) close to one end of the wire for  $\Delta_0 = \hbar \upsilon /R_0$ and $a_0 = 0.1 R_0$. Note that $|\Psi|^2$ peaks at the boundary between toroidal region and the disk-like end. It decays exponentially into the bulk of the wire (region $A$ as shown in Fig. (\ref{finite-wire3D})) and decays with a power law behaviour toward the centre of the disk (region $C$).} \label {Majowavefunc}
\end{figure}
The strategy for solving Eq.\ (\ref{zero})  is standard: we find the general solutions in the three regions $A$, $B$ and $C$, we match their wavefunctions at the boundaries and finally select the physical (normalizable) solution.  
The technical details are given in the Appendix B. Up to a normalization constant the solution is given by
\begin{equation}
\Psi(\zeta) =  \left \{ \begin{matrix}  \Psi_A (z)   &&   \zeta \in A \\  \\  \Psi_B (\theta)  &&    \zeta \in B \\ \\ \Psi_C(\rho) &&\zeta \in C     \end{matrix} \right .
\end{equation}
where the wave functions are defined as 
\begin{equation}\label{psiA}
\Psi_A (z) = \frac{1}{2}\left(\begin{matrix} 1 \\ -1 \\ 1 \\ -1 \end{matrix}\right) e^{\Delta_0 z}
\end{equation}
\begin{equation}\label{psiB}
\Psi_B (\theta) =\frac{1}{2} \left(\begin{matrix}  \cos{\theta \over 2} + \sin{\theta \over 2} \\  \\ \sin{\theta \over 2} - \cos{\theta \over 2} \\  \\ \cos{\theta \over 2} + \sin{\theta \over 2} \\ \\ \sin{\theta \over 2} - \cos{\theta \over 2} \end{matrix}\right) e^{\Delta_0 a_0 \theta}
\end{equation}
Assuming that the Zeeman term is negligible and considering only the solution which is well defined (i.e. no ambiguity in the phase at the apex which implies that the wave function should vanish at $\rho=0$) we obtain
\begin{equation}\label{psiC}
\Psi_C (\rho) = \sqrt{\rho \over 2 R_0} \exp{\left[\frac{1}{4}(\frac{\rho^2}{ R_0^2} - 1 + 2 \Delta_0 a_0 \pi )\right ]} \left( \begin{matrix} 1 \\ 0 \\  1  \\ 0 \end{matrix}\right)
\end{equation}

The probability density of the Majorana state is shown in Fig.\ (\ref{finite-wire3D}). Note that although the components of the wavefunction change in the $B$ region the absolute value follows the same behaviour as a function of length $\zeta$. The peak of the wavefunction is at the boundary between region $B$ and $C$. This way we obtain a solution to the Bogoliubov-de Gennes Hamiltonian which is real (up to an overall phase) and it is associated with a zero energy eigenvalue for a semi-infinite wire. The wavefunction is localized at the end of the wire in agreement with the numerical simulation and satisfies the Majorana condition $\hat{\Psi}_0^{\dag} = \hat{\Psi}_0$.

 \section{RESULTS ON STABILITY OF MAJORANA FERMIONS}
\subsection{Lattice model}
\begin{figure}[b]
\includegraphics[scale=0.33]{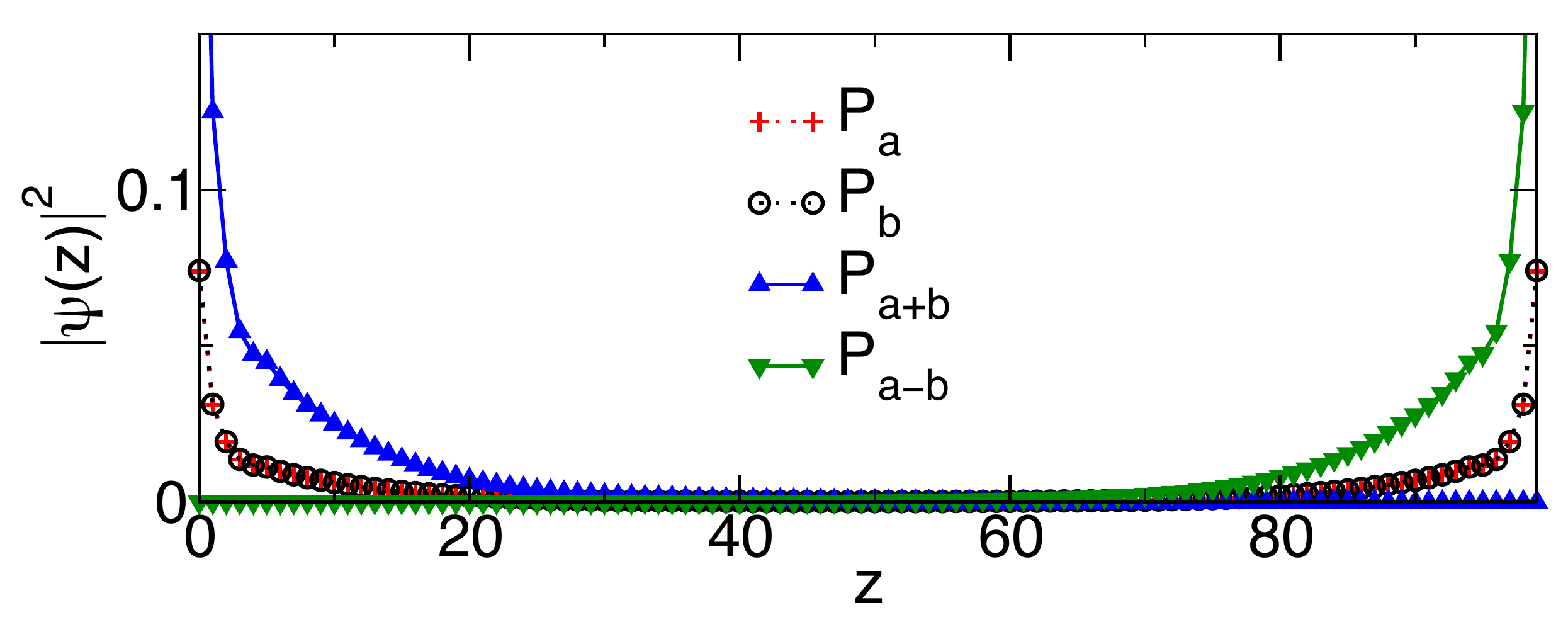}
\caption{Probability densities of Majorana end-states. $P_{a/b}(z)$
represents the particle component of the
wavefunction associated with $\pm E_0$
summed over $x$-$y$ coordinates. $P_{a\pm b}$ represent the even/odd superpositions of these wavefunctions.  The wire is 6 lattice sites wide in both the $\hat{x}$ and $\hat{y}$ directions and 100 lattice sites long in the $\hat{z}$ direction.  $\eta=0.49$ in units of the fundamental flux quantum, and the chemical potential $\mu=0.09\lambda$, where $\lambda=150$ meV.
}\label{probdens} 
\end{figure}
We establish the stability of Majorana fermions in the nanowire through a combination of additional analytical insights and numerical studies using the same concrete lattice model in \cite{cook} for the Bi$_2$Se$_3$ family
of materials \cite{qi_rev} given by 
Fu and Berg \cite{fu-berg1} regularized on a simple cubic lattice.  This model is defined by a
$k$-space Hamiltonian
\begin{equation}\label{latt1} 
h_\bk=M_\bk\sigma_1+\lambda\sigma_3(s_2\sin{k_x}-s_1\sin{k_y})+\lambda_z\sigma_2\sin{k_z},
\end{equation}
with $M_\bk=\epsilon-2t\sum_\alpha\cos{k_\alpha}$.  Here $\sigma_\alpha$
represent the Pauli matrices acting in the space of two
independent orbitals per lattice site. For $\lambda, \lambda_z>0$ and
$2t<\epsilon<6t$ the system described by Hamiltonian (\ref{latt1}) is
a TI in Z$_2$ class (1;000), i.e.\ a strong topological insulator. The
 magnetic field enters through the Peierls
substitution, replacing all hopping amplitudes as $t_{ij}\to
t_{ij}\exp{[-(2\pi i/\Phi_0)\int_i^j\bA\cdot d{\bf l}]}$ and the Zeeman
term $-g\mu_B \bB\cdot{\bs}/2$ where $\mu_B=e\hbar/2m_ec$ is the Bohr
magneton. In the SC state the BdG Hamiltonian takes the form of Eq.\
(\ref{hsc1}) with  $\Delta_\bk=\Delta_0is_2$ describing on-site
spin singlet pairing.

In the subsequent calculations we consider the above Hamiltonian on the real-space cubic lattice and in various wire geometries with rectangular cross sections and both periodic and open boundary conditions along the length of the wire. We find eigenstates and energy eigenvalues by the exact numerical diagonalization using standard LAPACK routines and by sparse matrix techniques in cases where only low-lying states are of interest. Unless explicitly stated otherwise we use the following set of model parameters in our subsequent calculations:  $\lambda_z=2\lambda$, $t=\lambda$, $\epsilon=4\lambda$. This places our model into Z$_2$ class (1;000) and with $\lambda=150$meV produces a bulk bandgap of 300meV, as in Bi$_2$Se$_3$ crystals.

As an example of a calculation based on the lattice model we show in Fig.\ \ref{probdens} the probablity density of Majorana end-states in a wire 100 lattice spacings long. We note that in any finite-length wire there will always be an exponentially small overlap between the two Majorana end-states. Such an overlap leads to the hybridization and a small non-zero energy $\delta E$ for the combined fermionic state which shows probablity density equally split between the two ends of the wire. An (unphysical) state with equal probablity density exists at energy $-\delta E$. Fig.\ \ref{probdens} shows how the Majorana end-states can be constructed by taking the appropriate linear superpositions of the above eigenstates. It is to be noted that for a finite-length wire the Majorana end-states are not true eigenstates of the system; they will mix on a time scale $\hbar/\delta E$ which is, however, exponentially long in the wire length $L$.

\subsection{The Majorana number and the topological phase diagram}
\begin{figure*} 
\includegraphics[scale=0.6]{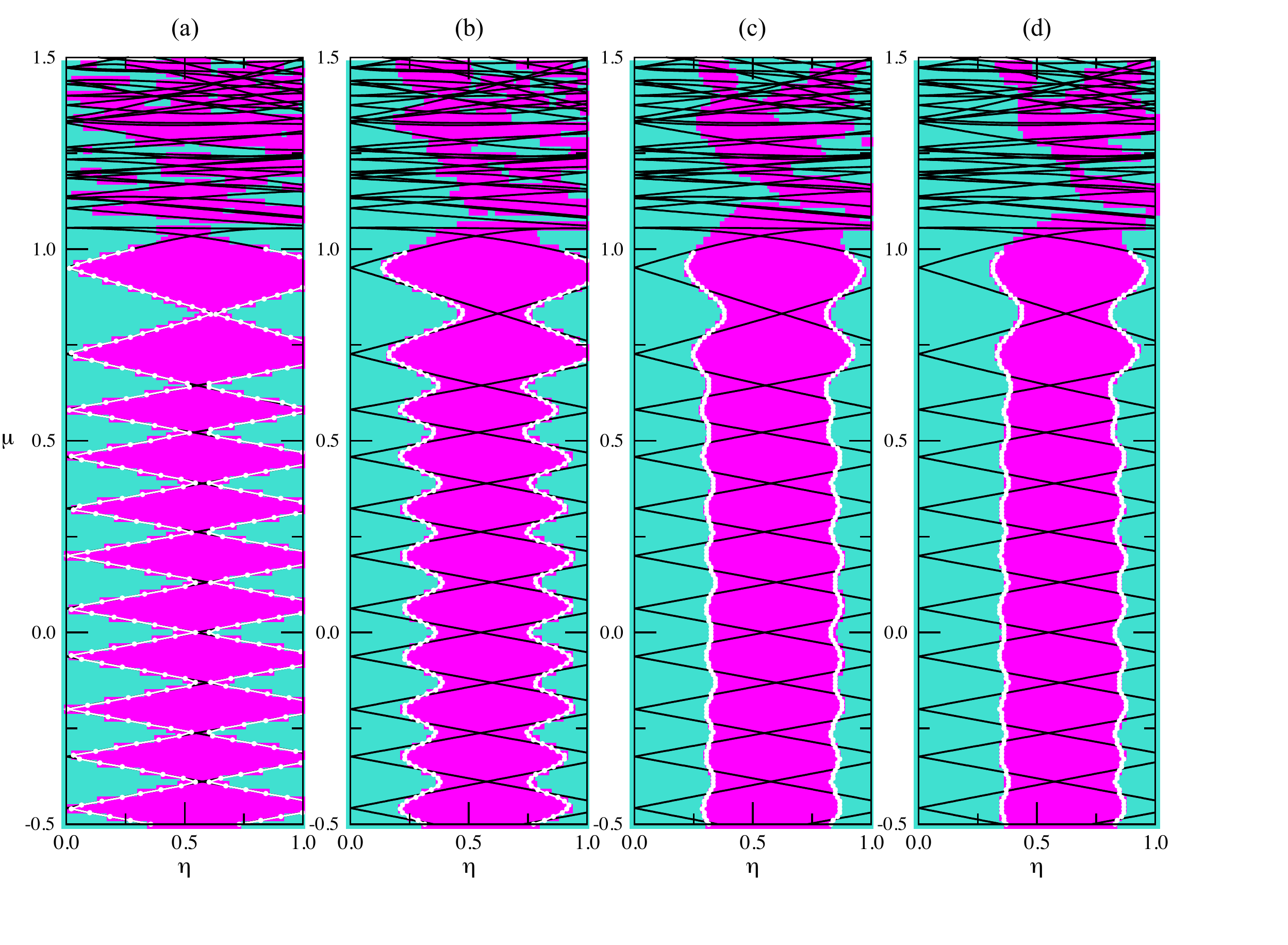} 
\caption{Phase diagrams computed from generalized Majorana number for the infinitely-long, clean wire with a 10 lattice site by 10 lattice site cross-section.  (a), (b), and (c) are each for a system with a vortex, while the system with phase diagram (d) lacks a vortex.  $|\Delta|$ is set to $0$ and $0.04\lambda$ in (a), (b), respectively, and $0.08\lambda$ in both (c) and (d).  Chemical potential $\mu$ is in units of $\lambda=150$ meV and $\eta$ is in units of the fundamental flux quantum.  Blue and pink regions were created by computing the generalized Majorana number $\cM$ in steps of at most $\Delta\eta=0.02$ and $\Delta\mu=0.02$, colouring these points blue ($\cM=1$, non-topological phase, no Majorana zero-modes in system) or pink ($\cM=-1$, topological phase, Majorana zero-modes present in system), respectively, and enlarging these data points to create regions of solid colour.  As well, the phase boundaries were computed (white lines) via a more efficient algorithm, with error bars at most the size of the symbols.  White lines extend only up to $\mu=1$ as at larger $\mu$ the topological phase regions break up into small domains and the algorithm used to compute the phase boundary is only effective for large, simply-connected regions of the phase diagram.
}\label{phasewithvortex}
\end{figure*}
The existence of MFs at the ends of the wire depends on whether or not the wire is in the topological phase. For a 1D system the presence of the topological phase is indicated by Kitaev's  Majorana number \cite{kitaev1} $\cM(H)$. In Ref.\ \cite{cook} we have computed $\cM(H)$ in the limit $\Delta\to 0$ where it reflects simply the parity of the number of the Fermi points of the underlying normal state in the right half of the Brillouin zone. The resulting phase diagram in the $\eta$-$\mu$ plane, for $\mu$ inside the bulk bandgap, consists of diamond-shaped topological regions indicated in Fig. \ref{phasewithvortex}(a). In the limit $\Delta\to 0$ the individual diamonds just touch at their apices yielding a continuous topological phase for a specific value of magnetic flux $\eta$ close to $1\over 2$ and for all values of $\mu$ inside the band gap. This feature underlies the key advantage of 
the present setup: the chemical potential does not require fine tuning. However, it is also true that when passing between individual diamond-shaped regions, the system gets arbitrarilly close to the phase boundary and one thus expect topological order to be rather fragile in these regions. On the other hand, away from the $\Delta\to 0$ limit one intuitively expects the topological phase to become more robust and indeed this was suggested by the numerical results presented in  Ref.\ \cite{cook}. In the following we shall elucidate this point and show that indeed for $\Delta_0>0$ the topological phase becomes a compact region in the $\eta$-$\mu$ plane.

We consider a general definition of the Majorana number\cite{kitaev1}
\begin{equation}\label{genmajnum} 
\cM(H)={\rm sgn}\left[{\rm Pf} \tilde{B}(k=0)\right]{\rm sgn}\left[{\rm Pf} \tilde{B}(k={\pi\over a})\right],
\end{equation}
where $\tilde{B}(k)$ is the position-space Hamiltonian of the infinite-length, lattice-model TI wire written in terms of Majorana fermion operators and Fourier-transformed in the $\hat{z}$-direction.  $a$ is the lattice-spacing, implying that $\tilde{B}(k)$ is evaluated here at $k=0$ and at the edge of the Brillouin zone.  Eq.\ (\ref{genmajnum}) simplifies to the previously-given definition in \cite{cook} when $|\Delta|$ is sufficiently small.

We consider the phase diagram in two limiting situations:  When $\Delta$ winds in phase counterclockwise by $2\pi$ around the circumference of the wire, corresponding to a vortex present along the length of the TI nanowire, and when $\Delta$ does not wind in phase, meaning there is no vortex in the system. These two situations are expected to represent the ground state of the system for the magnetic flux close to $\Phi_0/2$ and 0, respectively. The precise value of the flux at which the vortex enters will depend on details but we show below that, remarkably, the topological phase diagram is fairly insensitive to the presence or absence of the vortex.

Fig. \ref{phasewithvortex} displays the phase diagram of a rectangular wire with a $10\times 10$ cross section based on Eq.\ (\ref{genmajnum}).
As $|\Delta|$ is increased from zero, we see that the boundary of the region corresponding to the topological phase smoothes out, with the region centered close to $\eta=0.5$ and extending through all values of the chemical potential inside the bulk band gap and also up into the bulk conduction band. We remark that the topological phase here is centered near a value of the magnetic  flux that somewhat exceeds $\Phi_0/2$. This is because the surface state penetrates slightly into the bulk of the wire and thus encloses somewhat smaller amount of flux than the geometric surface area of the wire. For thicker wires this shift will be negligible.

 It is also interesting to note that according to Fig. \ref{phasewithvortex} the topological phase persists for $\mu$ well into the conduction band (as well as the valence band, which is not explicitly shown). This finding is potentially important in view of the fact that most TI crystals naturally grow with the chemical potential pinned inside the bulk conduction or valence band. We will show below, however, that although MFs indeed appear in this regime, the minigap protecting them quickly collapses as $\mu$ moves deeper into the bulk band. Heuristically, one can understand this as follows. With $\mu$ inside the conduction band the bulk of the nanowire becomes metallic. In the presence of SC order and with the magnetic field applied along its axis there will be a vortex line running along its center. Such a vortex line will host low-energy core states with the characteristic energy scale $\Delta^2/E_F$ which quickly becomes very small as $E_F$ increases (here $E_F$ is the Fermi energy measured relative to the bottom of the bulk conduction band). By contrast when the chemical potential lies inside the bandgap the bulk of the wire remains insulating and does not contribute any low-energy states.

From study of the phase diagram Fig.\ \ref{phasewithvortex} at $\Delta>0$, we begin to understand the robustness of the Majorana bound states. Consider first the effect of non-magnetic disorder, modeled by introducing a spatially fluctuating component of the chemical potential $\mu\to \bar{\mu}+\delta\mu(\br)$. Assume also that $\delta\mu(\br)$ is slowly varying in space so that only variation along the $z$-direction are meaningful and $\mu(z)$ can be thought of as defining a phase of the wire in the vicinity of the coordinate $z$. 
If the average chemical potential $\bar{\mu}$ and  the flux $\eta$ are such that the system starts deep inside the topological phase then it is clear that fluctuations in $\delta\mu$ will not drive the system out of the topological phase unless they exceed the bulk bandgap. 
Similarly, fluctuations in the total magnetic flux $\eta$, which can occur e.g.\ in a wire with a non-uniform cross section, will not drive the system out of the topological phase unless they reach a significant fraction of $\Phi_0/2$. We demonstrate below by explicit inclusion of disorder in the lattice model that the heuristic argument given above remains valid even when disorder potential varies rapidly on the lattice scale.

The smoothing out of the topological region's boundary as $|\Delta|$ is increased can be understood by studying the low-energy analytical theory again.  We start from the Hamiltonian in Eq. \ (\ref{hsc2}) and let $\mu=m=0$, studying how the phase diagram changes for this value of the chemical potential as $|\Delta|$ is increased from zero.  If we now assume that $\eta$ deviates from 1/2 by a small amount, i.e.\ $\eta=1/2+\delta\eta$ then the spectrum for the $l=0$ branch can be written as
\begin{equation}\label{Ek_etanonzero2} 
E_k = \pm\left[k^2+\left({{\delta\eta}\over R}\pm\Delta_0\right)^2\right]^{1/2}. 
\end{equation}
We know that the system will be in the topological phase when $\delta\eta=0$ and $\mu=0^\pm$. To understand the smoothing out of the topological region's boundary, we identify when the spectral gap closes for non-zero $\Delta_0$ as a function of $\delta\eta$, since closing of the gap signifies a phase transition.  
We notice that the gap in Eq.\ (\ref{Ek_etanonzero2}) remains as $\delta\eta$ is moved away from $0$ until $\delta\eta=\pm R\Delta_0/v\hbar$, where we have restored proper units.  Therefore, we see that at $\mu=0$, for nonzero $\Delta_0$, the topological phase has widened from a point at $\eta=1/2$ to an interval $\left(1/2-R\Delta_0/v\hbar,1/2+R\Delta_0/v\hbar\right)$, as observed in the phase diagrams computed numerically using the more general definition of the Majorana number. 
 
\begin{figure}[t]
\includegraphics[width = 8.9cm]{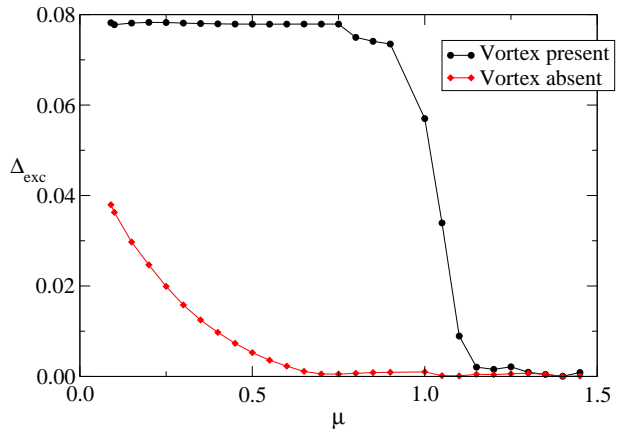}
\caption{Quasiparticle excitation gap $\Delta_{\rm exc}$ of an infinitely-long wire without disorder with a 14 lattice site by 14 lattice site cross-section as a function of chemical potential $\mu$ with vortex present (black circles) and vortex absent (red diamonds). $\Delta_{\rm exc}$ and $\mu$ are expressed in units of $\lambda=150$ meV.  Here $|\Delta|=0.08\lambda$.
}\label{SCgap}
\end{figure}

The absence or presence of a vortex in the TI nanowire makes negligible differences to the phase diagrams as seen by comparing Figs.\ \ref{phasewithvortex}(c) and \ref{phasewithvortex}(d).  However, the presence or absence of a vortex does have a pronounced effect on the quasiparticle excitation gap $\Delta_{\rm exc}$ (shown in Fig. \ref{SCgap}) in the infinitely-long wire.   
From Fig. \ref{SCgap}, we see that $\Delta_{\rm exc}$ remains close in magnitude to $|\Delta|$ up until $\mu$ reaches the bottom of the bulk conduction band if the SC order parameter winds counter-clockwise in phase by $2\pi$, while without a vortex $\Delta_{\rm exc}$ can be seen to quickly decay with increasing $\mu$. Clearly, near $\eta=1/2$ the former represents a more physical situation. 

\begin{figure}[b]
\includegraphics[width = 8.9cm]{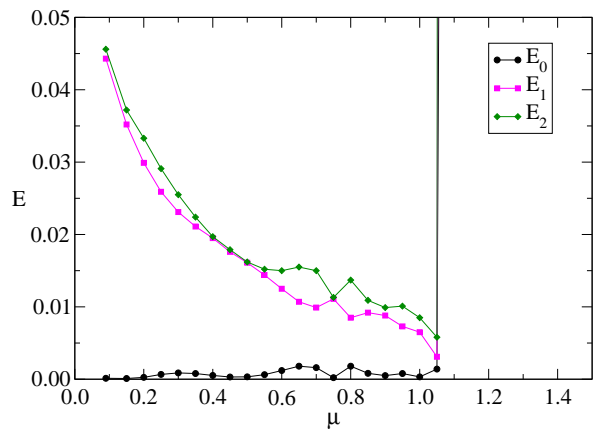}
\caption{Three lowest energy eigenvalues $E_2$, $E_1$, and $E_0$ of the finite-length wire, without disorder, with 14 by 14 by 100 lattice sites as a function of chemical potential $\mu$ with vortex present with $|\Delta|=0.08\lambda$. $\Delta_{exc}$ and $\mu$ are expressed in units of $\lambda=150$ meV. Eigenvalues were computed via Lanczos method and failed to converge for $\mu>1.05$, resulting in a non-physical spike to the non-convergent next data point.  
}\label{minigap_vortex}
\end{figure}

For experimental realization of the device, it is important to know how large the minigap - the energy of the lowest non-zero mode coming from bound states at the ends of the wire which are absent when the wire is infinitely long - is as well as studying the quasiparticle excitation gap, which is the energy of the lowest non-zero mode for the infinitely-long wire.  If we study this minigap as a function of chemical potential with a vortex present, shown in Fig. \ref{minigap_vortex}, we see that, for $\mu$ close to zero, the minigap starts with an amplitude close to the SC gap $\Delta$, in accord with the analytical theory presented in Sec. III.B. The minigap then continuously decreases with increasing $\mu$, retaining a respectable value $\sim 0.1\Delta$ at the edge of the bulk conduction band at $\mu=1$. As mentioned previously, the minigap then quickly collapses as the bulk bands are populated but nevertheless persists over a non-zero range of $\mu>1$.  
\subsection{Robustness of Majorana fermions against disorder}
As mentioned previously, we expect robustness of the Majorana end states  with respect to non-magnetic disorder in the proposed device. There are two related but logically separate issues that pertain to this problem. First, we must ensure that SC order induced in the wire is itself robust against non-magnetic disorder. Since the device can be operated at (or very close to) the time-reversal invariant point and since we consider a spin-singlet s-wave SC order we expect this to be the case on the basis of Anderson's theorem. Below, we illustrate this aspect of the robustness by performing a self-consistent numerical calculation on our model in the presence of disorder. Second, we must show that Majorana end-states themselves remain stable in the presence of disorder. To some extent this already follows from our arguments in the previous subsection based on the study of the topological phase diagram. Also, stability of MFs follows from the stability of the SC phase in the bulk of the wire argued above. Nevertheless, to address this question directly, we perform explicit numerical calculations for open-ended wires in the presence of non-magnetic disorder and in various physical regimes. These calculations confirm the expected robustness of MFs and yield additional insights into the quantitave aspects of this robustness; specifically they provide information about the minigap magnitude and the mechanism by which MFs are eventually destroyed in the strong-disorder limit.    

To study these questions, we add a term corresponding to disorder in the on-site potential to the lattice Hamiltonian, $H_0$, of the clean system.  The Hamiltonian for the system with disorder, $H_{\rm dis}$, is therefore
\begin{equation}\label{Hdis} 
H_{\rm dis}=H_0+\sum_{i\alpha}U_ic^\dagger_{i\alpha}c_{i\alpha'},
\end{equation}
where $U_i$, the random  potential at lattice site $i$, is assigned a value from a uniform, random distribution ranging from ${-U\over2}$ to ${U\over2}$.

\begin{figure}[t]
\includegraphics[width=8.9cm]{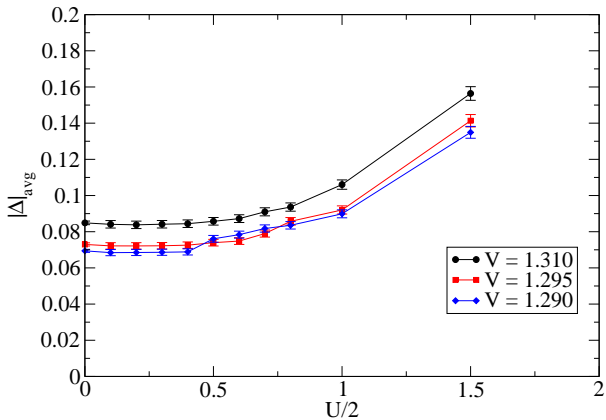}
\caption{Mean superconducting order parameter magnitude $|\Delta|_{\rm avg}$ for three different values of the pairing potential ($V=1.310$, $V=1.295$, and $V=1.290$) as a function of disorder strength $U\over2$.  $|\Delta|_{\rm avg}$ is averaged over every lattice site in a 6x6x6 lattice site nanowire with periodic boundary conditions in the $\hat{z}$-direction with random disorder in the chemical potential and also further averaged over 10 such randomly-disordered nanowires.  The mean chemical potential for all data points is $\mu_{\rm avg}=0.09\lambda$.  The self-consistent calculation for each disordered nanowire proceeded until the maximum difference in the superconducting order parameter magnitude between the final iteration and the next-to-last iteration of the calculation at any lattice site was $0.001\lambda$.  
}\label{selfconsistentdel}
\end{figure}

As a first step we compute the magnitude of the SC order parameter self-consistently as described in \cite{tinkham1} for different disorder strengths and with periodic boundary conditions along $z$, i.e.\ no Majorana end-states. This calculation assumes the existence of an intrinsic pairing potential $V$ in the wire and is therefore, strictly speaking, not directly relevant to the proximity-induced SC state discussed in the rest of this paper. It  nevertheless illustrates very nicely the robustness of the SC order with respect to disorder. The key point is that one would expect SC order to be even {\em more} robust when induced by proximity to the bulk superconductor.   
Fig.\ \ref{selfconsistentdel} shows mean SC order parameter magnitude vs. $U\over2$ computed for different values of the pairing potential $V$.  We observe that $|\Delta|_{\rm avg}$ first decreases slightly with increasing $U$, but at larger $U$ the mean SC order parameter increases in magnitude.  We attribute this increase to the buildup of the normal density of states at the Fermi level, $N(\mu)$, due to disorder. Such a buildup is known to occur in other 2D systems with a Dirac spectrum\cite{neto77} and it should increase $|\Delta|$ according to the standard BCS formula\cite{tinkham1}
\begin{equation}\label{Hdis} 
\Delta=\hbar\omega_c e^{-1/{N(\mu)V}}.
\end{equation}
Here $\omega_c$ and $V$ are constants.  These results suggest the SC order parameter is not only quite robust against non-magnetic disorder but the latter can acually enhance it when the chemical potential is close to the Dirac point.

\begin{figure}[t]
\includegraphics[scale=0.7]{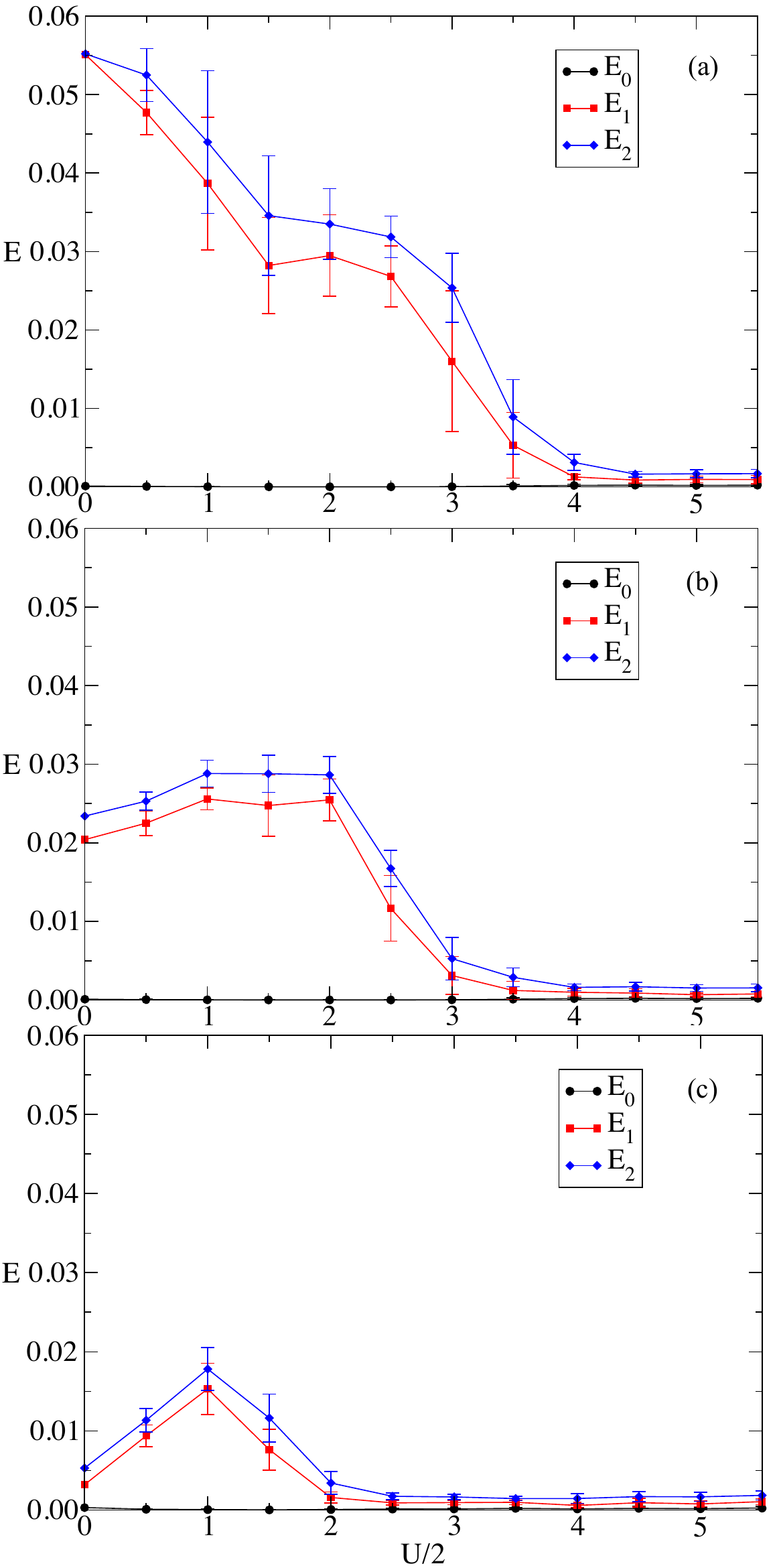}
\caption{Three lowest positive energy eigenvalues $E_0$, $E_1$, and $E_2$ of a finite-length TI nanowire model with 6 by 6 by 100 lattice sites obtained by the Lanczos method as a function of disorder strength
 ${U\over2}$ for mean chemical potential values of $\mu=0.09$ (a), $\mu=0.4$ (b), and $\mu=0.8\lambda$.  $|\Delta|=0.08\lambda$ for each. ${U\over 2}$ and energy $E$ are expressed in units of $\lambda=150$ meV.  The error bars reflect averaging over 10 independent realizations of the random potential.
}\label{disorderSCgap_mu0_09}
\end{figure}

To address the robustness of Majorana end-states we now study the finite-length wires.
Using sparse matrix techniques, we solved for the average values of the three lowest, positive eigenvalues of the spectrum, and plotted these for a range of $U$, as shown in Fig. \ref{disorderSCgap_mu0_09}. These calculations are performed for a constant value of the SC gap, having previously established that disorder has only a weak effect on the latter.  We see that the energy of the  Majorana bound state remains very close to zero, with no observable fluctuations.  The topological SC is eventually destroyed by the collapse of the minigap, i.e. lowering of the excited states at some critical value of disorder strength $U_c$.  It is interesting to note that $U_c$ is rather large, being expressed in units of $\lambda=150$ meV, exceeding the TI bulk bandgap by more than a factor of 5.  Furthermore, the MFs are robust against disorder at a wide range of average chemical potential in the bulk band gap, although the minigap is largest for values close to the middle of the bulk band gap and smallest for values near the edge of the bulk band gap.  We note that the minigap is roughly the same for mean chemical potential values of $0.09\lambda$, $0.4\lambda$, and $0.8\lambda$ if ${U\over2}=\lambda$, suggesting disorder stabilizes the minigap in this regime. It is also interesting to note that for two larger values of $\mu$ disorder initially increases the minigap thus making the topological phase more robust.

\section{Conclusion and Discussion}
\label{disc}
The main goal of this work was to study the stability of Majorana zero-modes in the proposal introduced in \cite{cook}, which consists of a TI nanowire, proximity-coupled to a bulk s-wave superconductor, with a weak magnetic field applied along the nanowire's length.  After reviewing the literature to illustrate the importance of identifying a device in which MFs emerge under a wide range of accessible conditions, and reviewing the theory behind the proposed device in greater detail than possible in \cite{cook}, we studied the robustness of the topological superconductor phase of the device against disorder.

As a first step we computed the topological phase diagram of the TI nanowire numerically in a semi-realistic lattice model. Using a general definition of Kitaev's Majorana number $\cM(H)$ we were able to show that for non-zero values of SC order parameter $\Delta$ the topological phase forms a set of compact columnar regions in the plane spanned by the magnetic flux $\eta=\Phi/\Phi_0$ and the chemical potential $\mu$, centered around half-integer values of $\nu$ and covering about $50\%$ of the phase diagram (see Fig.\ \ref{phasewithvortex}). This form of the phase diagram confers two principal advantages of our proposed device as regards future experimental realizations and potential applications: (i) unlike in the semiconductor wire 
realizations\cite{be150} where significant fine-tuning of the chemical potential is required to reach the topological phase, our proposed device produces Majorana end-states for $\mu$ anywhere inside the bulk bandgap of the TI; and (ii) if the average chemical potential of the nanowire is in the bulk band gap, we can expect the entire nanowire to remain in the topological phase even for large disorder strengths, as even then any local chemical potential value will remain in the topological region of the phase diagram.  We also observe from the phase diagram  Fig.\ \ref{phasewithvortex} that the topological phase persists over a wide range in magnetic flux through the cross-section of the nanowire. This feature is less critical because the magnetic field can be easily tuned in a laboratory.  Nevertheless understanding the magnetic phase boundary is very useful since changing the magnetic field strength can be used to tune the wire between topological and normal phases.

To explicitly ascertain the robustness of the MF with respect to thermal fluctuations and non-magnetic disorder we studied the system's minigap by both analytic and numerical techniques. Minigap, defined as the smallest non-zero energy level in the system, can be taken as a good proxy for the robustness of the quantum information encoded in MFs since uncontrolled excitations of quasiparticles out of the ground state into the low-lying excited states would obviously spoil such encoding. Also, large values of the minigap can aid experimental detection of the Majorana zero mode using various spectroscopic techniques.
We use the low-energy, analytical theory of \cite{cook} to first show that excited states should be close in energy to the magnitude of the superconducting gap when the chemical potential is close to the middle of the bulk band gap of the TI nanowire, indicating Majorana zero-modes in the device should be protected by a sizeable minigap.  We then employ the lattice model and compute the three lowest eigenenergies of the TI nanowire with disorder to show that the minigap remains significant even for the disorder strength considerably exceeding the bulk band gap of the TI.  Interestingly, we find that disorder strength comparable to the magnitude of the bulk bandgap also appears to stabilize the minigap at a sizeable value over changes in the average chemical potential in the nanowire, which might be useful in future applications of the device.

Stability of MFs in our proposed device also follows from the general
periodic classification of topological insulators and superconductors
given by  Schnyder et al.\cite{schnyder33} and by Kitaev\cite{kitaev33}. According to this classification Majorana bound
states may appear in one-dimensional systems of the D and DIII
symmetry classes. The former corresponds to the superconducting state
with broken time-reversal and spin symmetries while the latter is a
superconductor with only spin symmetry broken. We found that MFs are
most stable in our device with exactly half flux quantum,
corresponding to $\cT$-preserving DIII symmetry class. In this class
all the states are doubly degenerate according to the Kramers
theorem. It is thus slightly counterintuitive that we obtain isolated
non-degenerate MFs in this situation. To clarify this point we note
that the device will be in the DIII class only when the wire is
considered infinitely long (or else periodic along the
$z$-direction). MFs on the other hand occur only in a finite wire
where $\cT$ is explicitly broken at the wire ends by the
magnetic flux lines entering and exiting the wire.

The above considerations also suggest stability of the MFs with
respect to magnetic disorder which was not explicitly studied in this
paper.  For one, magnetic impurities are much less dangerous for
proximity-induced SC than intrinsic SC.  Furthermore, magnetic
impurities will not further violate the class D symmetry possessed by
this system when $\cT$ is broken.

\begin{figure}[t]
\includegraphics[width = 8.9cm]{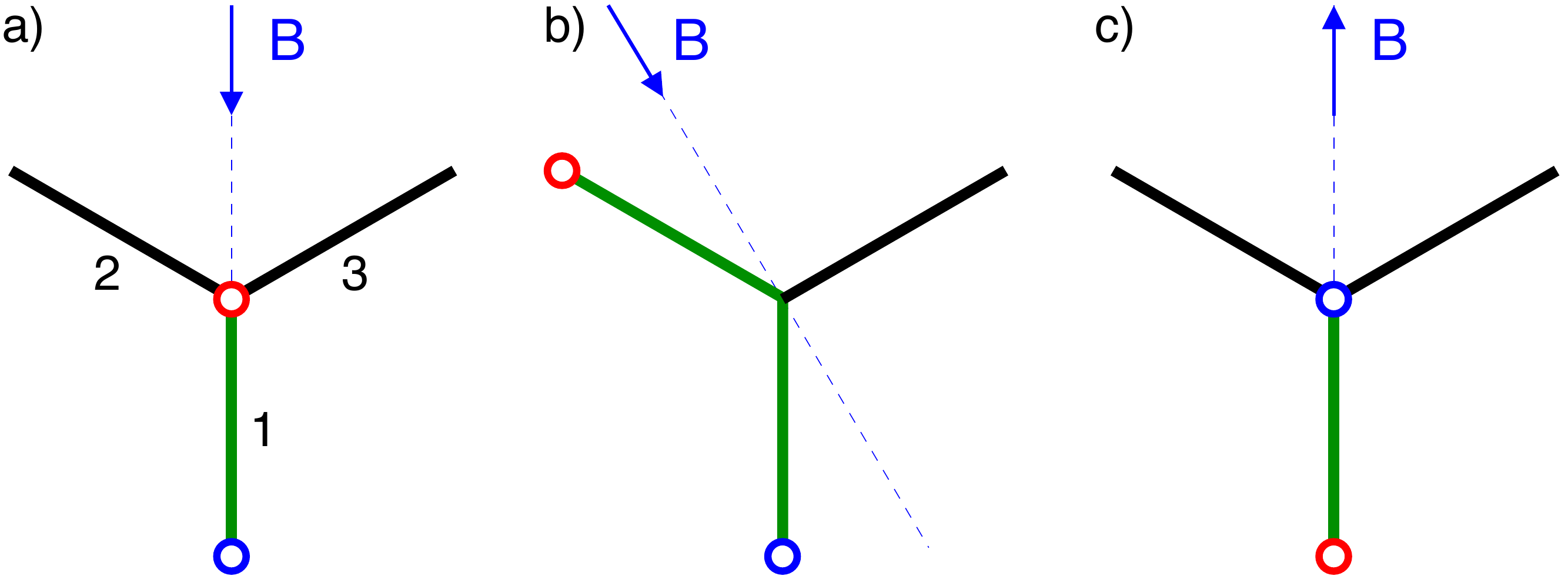}
\caption{Exchange of MFs in a trijunction device. a) The field $\bB$ is tuned so that nanowire 1 has flux close to $\Phi_0/2$ through its cross section and is thus in the topological phase with MFs localized near its ends. The flux through wires 2 and 3 is down by the factor $\cos(2\pi/6)=0.5$ and they are thus in the trivial phase. b) Rotating the direction of $\bB$ by $30^{\rm o}$ the flux through wires 1 and 2 becomes $\cos(2\pi/12)(\Phi_0/2)\simeq 0.866 (\Phi_0/2)$ and is thus sufficiently close to $\Phi_0/2$ for them both to be in the topological phase according to the phase diagram in Fig.\ \ref{phasewithvortex}. As a result the MF previously located at the junction (red circle)  has now moved to the other end of wire 2 as indicated. Continuing this process by rotating $\bB$ further in $30^{\rm o}$ increments it is easy to map out the motion of MFs and conclude that after $180^{\rm o}$ rotation the system comes back to the original situation with MFs localized on wire 1 but with their order exchanged as illustrated in panel c).
}\label{trijunction}
\end{figure}
Before concluding we address the following question: Do MFs predicted to exist in our device obey non-Abelian exchage statistics? This property is obviously of paramount importance for any future application in quantum information processing. Alicea et al.\ \cite{alicea77} clarified the sense in which MFs in 1D wire networks exhibit non-Abelian statistics upon exchange, considering semiconductor wires\cite{sau2,oreg1,be150}, and showed how particle exchanges can be effected in such a setting. Although superficially similar to these models\cite{sau2,oreg1} our proposal is more closely related to the original Fu-Kane vortex at the interface between a TI and an ordinary superconductor\cite{a49}. Indeed consider a thought experiment in which we take our wire and slowly increase its radius while simultaneously reducing the applied magnetic field so that the total flux through its cross section remains constant at $\Phi_0/2$. We also assume that all the surfaces of the resulting disk remain covered by a thin SC film. In the limit when the radius $R$ becomes comparable to the wire length $L$ we have a bulk disk-shaped TI covered with a SC film. The presence of $\Phi_0/2$
flux and the cylindrical symmetry dictates that a vortex must be present at the center of each of the flat disk surfaces. Such vortices will contain MFs\cite{a49} and will obey non-Abelian exchange statistics according to the standard arguments. MFs in our wires are thus adiabatically connected to those residing in the cores of Fu-Kane vortices and are therefore expected to obey the same non-Abelian exchange statistics when organized into T-junctions or wire network geometries\cite{alicea77}. In Fig.\ \ref{trijunction} we outline a simple protocol that implements an exchange of two MFs in a symmetric `trijunction' device formed by three nanowires joined at a single point. The exchange of two MFs, initially localized at the ends of one of the wires, is effected simply by a $180^{\rm o}$ rotation of the applied magnetic field in the plane of the device.

We conclude that unpaired Majorana zero-modes are exceptionally stable in our proposed device, being present over a wide range in magnetic flux, chemical potential, and disorder strength, with disorder even being expected to stabilize the MFs to an extent. They obey non-Abelian exchange statistics by virtue of being adiabatically connected to the Fu-Kane vortex\cite{a49} and, as illustrated above, can be easily manipulated by changing the direction of the applied magnetic field. On this basis we expect the device architecture discussed in this work to be useful for future experimental study of MFs and their their potential applications. 

\section{Acknowledgements}

We thank J. Alicea, S. Frolov, L. Fu, C. Kane, A. Kitaev, R. Lutchyn, G. Refael and X.-L. Qi for valuable comments and correspondence. The work reported here was supported by NSERC and CIfAR.

\appendix

\section{Surface state in a thin wire}

To study the surface state in a thin wire we employ a continuum version of the 3D lattice model defined in Eq.\ (\ref{latt1}). Focusing on the vicinity of the $\Gamma$-point we thus write the following $4\times 4$ matrix Hamiltonian,
\begin{equation}\label{hh1}
h_k=-i\sigma_3(s_2\pi_x-s_1\pi_y)+\lambda_z\sigma_2k_z+\sigma_1m(r),
\end{equation}
where $\pi_j=-i\partial_j-A_j$ is the gauge-invariant momentum operator and $m(r)$ is the gap at the $\Gamma$-point which we take to have a radial depence to model an interface between a TI ($m>0$) and an ordinary insulator or vacuum ($m<0$). We also assume traslational invariance along the wire, thus keeping the $k_z$ quantum number (which we denote as $k$ hereafter).

We start by considering a $\delta$-function flux $\bA=\eta\Phi_0(\hat{z}\times\hat{\br})/2\pi r$ for which we can find an analytic solution. Transforming into the cylindrical coordinates we obtain
\begin{equation}\label{pi}
s_2\pi_x-s_1\pi_y=
\begin{pmatrix} 0 & e^{-i\varphi}(-\partial_r+{\eta\over r}+{i\over r}\partial_\varphi) \\
 e^{i\varphi}(\partial_r+{\eta\over r}+{i\over r}\partial_\varphi) & 0
\end{pmatrix}.\nonumber
\end{equation}
We now perform a unitary transformation in the spin space defined by the matrix 
\begin{equation}\label{uni}
U_l(\varphi)=
\begin{pmatrix} e^{i\varphi l} & 0 \\
0 & e^{i\varphi(l+1)}
\end{pmatrix}.
\end{equation}
with $l$ the integer angular momentum quantum number. This leads to a decoupling in the angular variable with the individual channels described by
\begin{eqnarray}\label{hh2}
h_{kl}(r)&=& -\sigma_3\left[is_2\left(\partial_r +{1\over 2r}\right) + s_1{1\over r}\left(l+{1\over 2}-\eta\right)\right] \nonumber \\
&+&\sigma_2\lambda_zk+\sigma_1m(r).
\end{eqnarray}

From our previous discussion of the surface state we expect the gapless mode to occur for $l=0$ and $\eta={1\over 2}$. Furthermore, the band crossing occurs at $k=0$ so we focus on this value of $k$ and study 
\begin{equation}\label{hh3}
h_{00}(r)= -i\sigma_3s_2\left(\partial_r +{1\over 2r}\right)
+\sigma_1m(r).
\end{equation}
By performing a $\pi$-rotation around $\sigma_1$ and around $s_1$ this Hamiltonian can be brought into an off-diagonal form, 
\begin{equation}\label{hh3}
h_{00}(r)= 
\begin{pmatrix}
0 & -s_3(\partial_r +{1\over 2r}) +m(r) \\
s_3(\partial_r +{1\over 2r}) +m(r) & 0
\end{pmatrix},
\end{equation}
suitable for identifying the possible zero modes. We now describe the TI wire of radius $R$ by the following configuration of the gap function,
\begin{equation}
m(r) =  \left \{ \begin{matrix}  m_0   &&    r<R, \\
-m_0'  &&   r>R,      \end{matrix} \right .
\end{equation}
with $m_0$, $m_0'$ positive constants. For a wire placed in vacuum we take $m_0'\to \infty$. With this gap configuration it is easy to see that exactly two zero modes exist,
\begin{equation}\label{psi}
\psi_\alpha(r)=\chi_\alpha C\sqrt{R\over r}e^{m_0 r}, \ \ \ r<R, \ \ \ \alpha=1,2,
\end{equation}
with $C=\sqrt{2m_0/(e^{2m_0R}-1)R}$ the normalization constant and $\chi_1=(0,1,0,0)^T$, $\chi_2=(0,0,1,0)^T$. Thus, we conclude that exact zero modes exist for any radius $R$ in the presence of a $\delta$-function flux. This implies that the gapless surface modes also persist in this case even for a thin wire. We note however that only when  $R\gg\zeta=1/m_0$ are the zero-mode wavefunctions localized near the surface; for a thin wire they  permeate the entire bulk of the wire.

The above result can be understood based on a simple general argument. With the $\delta$-function half-flux the 3D Hamiltonian (\ref{hh1}) remains $\cT$-invariant. Therefore, Kramers theorem protects the band crossing at $k=0$,  and the surface mode must remain gapless for arbitrary $R$. This understanding suggests that the degeneracy at $k=0$ will be split by a more general flux distribution (e.g.\ that corresponding to a uniform $B$ field), and the size of the gap will then reflect the `amount' of $\cT$-breaking present in the system. Unfortunately, we were unable to find an exact solution for a more generic flux distribution. 

To test the above hypothesis we therefore proceed as follows. We start from the exact solution (\ref{psi}) of the Hamiltonian $h_{00}$ and add to it magnetic field $\delta \bB(r)$ as a perturbation. For simplicity we take the form $\delta \bB(r)=\hat{z}(ar-b)$ with constants $a$ and $b$ chosen so that the total additional flux $\delta\Phi$ vanishes. The corresponding vector potential is of the form
\begin{equation}\label{da}
\delta\bA= {\eta\Phi_0\over \pi R^2}(\hat{z}\times \br)\left({r\over R}-1\right),
\end{equation}
where the overall prefactor has been chosen so that $\delta B(R)=\eta\Phi_0/\pi R^2$, i.e.\ the field strength at the surface of the cylinder is the same as it would be in the case of a uniform magnetic field.

Inclusion of the above vector potential leads to the following additional term in the Hamiltonian, 
\begin{equation}\label{dh}
\delta h=\sigma_3 s_1 2\eta{r\over R^2}\left({r\over R}-1\right),
\end{equation}
which we treat in the first order perturbation theory. It is straightforward to evaluate the requisite matrix element $\langle\psi_1|\delta h |\psi_2\rangle$. For $\eta={1\over 2}$ and in the limit of $\zeta\ll R$ this leads to a correction to the energy of the form 
\begin{equation}\label{de}
\delta E=\pm{1\over 2} m_0\left({\zeta\over R}\right)^2.
\end{equation}
The energy splitting is seen to be proportional to the ratio between the cross-sectional areas of the surface wavefunction and of the cylinder. Thus, we arrive at the conclusion that the gap in the spectrum is indeed proportional to the amount of the $\cT$-breaking in the system as measured by the exposure of the surface wavefunction to the magnetic field.

\section{Dirac Hamiltonian for a finite wire configuration}
Any cylindrically shaped surface with slight rotationally invariant deformations from ideal cylinder can be described by the function $R=R(z)$ in which $z$ is the distance from an arbitrarily chosen origin on the $z$ axis (the axis of the deformed cylinder) and $R(z)-R_0$ is the deviation from the ideal cylinder with radius $R_0$. One can use Eq.(\ref{h01}) to find the Hamiltonian for the surface electrons of a TI wire with such a configuration if $R(z)$ and its first derivative is continuous. In order to do so it is convenient to use a parameter, $\beta$, which is defined as the angle between the normal vector $\hat{\bn}$ and the plane perpendicular to the axis of the wire. Note that we have assumed that the deformation from the ideal cylinder is such that the axis of the initial ideal cylinder always remains inside the wire and there is a rotational symmetry around that axis. This way we can write the normal unit vector in the cylindrical coordinate system
\begin{equation}
\hat{\bm n}(\varphi,z) = \cos{\beta(z)} \hat{\bm \rho}(\varphi) + \sin{\beta(z)} \hat{z} 
\end{equation}
for the ideal cylinder $\beta(z) = 0$ (region $A$ as shown in Fig. (\ref{finite-wire3D})). $\beta$ gradually goes to $\pm \pi/2$ at the smooth ends of the wire. 
Using Eq. (\ref{h01}) we can obtain the continuum real space representation of the $2\times 2$ Hamiltonian matrix for the surface states in the presence of the magnetic flux 
\begin{equation}
\frac{1}{\hbar \upsilon} h_0 =  \mu_{\text{eff}}(z) + s_{\varphi} \text{D}_1(z)  + s_{\rho} \text{D}_2(\varphi,z)  + s_3 \text{D}_3(\varphi,z)
\end{equation}
in which $s_{\varphi} = - s_1 \sin{\varphi} + s_2 \cos{\varphi}$ and $s_{\rho} = s_1 \cos{\varphi} + s_2 \sin{\varphi}$ and the functions above are defined as
\begin{equation}
\mu_{\text{eff}}(z) = \frac{1}{2} \left (\frac{1}{R(z)} + \frac{d \beta(z)}{d z} \right) \cos{\beta(z)} 
\end{equation}
\begin{equation}
\text{D}_1(z) = i \cos{\beta(z)}\partial_z  + \frac{1}{2} \sin{\beta(z)} \left(\frac{1}{R(z)} - \frac{d \beta(z)}{d z}\right)
\end{equation} 
\begin{equation}
\text{D}_2(\varphi,z) =  \frac{\sin{\beta(z)}}{R(z)} \left(i \partial_{\varphi} + \eta(z) \right)
\end{equation} 
\begin{equation}
\text{D}_3(\varphi,z) =  -\frac{\cos{\beta(z)}}{R(z)} \left(i \partial_{\varphi} + \eta(z) \right) - \frac{g_s \mu_B}{\hbar \upsilon} B_0
\end{equation} 
$\eta(z)$ is the fraction of the magnetic flux in the units of $h/e$ that is enclosed by the radius $R(z)$. 

The same unitary transformation that has been used in defining the spinors given in Eq. (\ref{psi1}) can transform $(s_{\rho},s_{\varphi})$ to $(s_1, s_2)$ and therefore it can make the new Hamiltonian matrix invariant under rotation $\varphi \rightarrow \varphi + 2 \pi$ 
\begin{equation}
U(\varphi) = \left(\begin{matrix} 1 && 0 \\  \\ 0 && e^{ i \varphi} \end{matrix}\right)
\end{equation}
One has to be careful with the partial derivative with respect to $\varphi$ since it does not commute with the transformation matrix $U(\varphi)$. It is easy to check that it transforms as follows
\begin{equation}\label{U01}
i U^{\dag}(\varphi)  \partial_{\varphi} U(\varphi)  = \frac{s_3}{2} + i \partial_{\varphi}
\end{equation}
This way the transformed Hamiltonian is invariant under rotations about the axis of the wire with the associated quantum number $l$. Therefore, similar to our treatment of the infinite wire, for each $l=0,\pm 1,\dots$ we obtain
\begin{equation}
\frac{1}{\hbar \upsilon} \tilde{h}_{l}(z) =  \tilde{\mu}_{\text{eff}}(z) + s_2 \text{D}(z)  + s_1 \text{m}_{1 l}(z)  + s_3 \text{m}_{2 l}(z)
\end{equation}
The functions used in the above expression are defined as
\begin{equation}
\tilde{\mu}_{\text{eff}}(z) = \frac{1}{2} \cos{\beta} \frac{d \beta}{d z}
\end{equation}
\begin{equation}
\hat{D}(z) = \frac{i}{2} ( \partial_z \cos{\beta} + \cos{\beta}  \partial_z)
\end{equation}
\begin{equation}
\text{m}_{1 l}(z) = - \frac{\sin{\beta}}{R(z)} (l + \frac{1}{2} - \eta(z))
\end{equation}
\begin{equation}
\text{m}_{2 l}(z) =  \frac{\cos{\beta}}{R(z)} (l + \frac{1}{2} - \eta(z)) + m_{\rm Zeeman}
\end{equation}

One can recover the infinite cylinder Hamiltonian given in Eq. (\ref{dir2}) by replacing $R(z)$ with $R_0$ and putting $\beta$ to zero. Therefore in the region $A$ and for $\eta = 0.5$ and $l=0$ we get
\begin{equation}
h_A (z) =  i s_2  \partial_z
\end{equation}
The region $B$ Hamiltonian can be obtained similarly. Here we have $\beta = \theta$ where $\theta$ is the angle that parametrizes the quarter circle cross the section of the surface in region $B$. It varies from zero at the $A/B$ boundary to $\pi/2$ at the $B/C$ boundary. We thus have
\begin{equation}
R(z) = R_0 - a_0(1 - \cos{\theta})
\end{equation}
\begin{equation}
z = a_0 \sin{\theta}
\end{equation}
Assuming that $a_0 \ll R_0$ and using $\theta$ as our new coordinate we can obtain the Hamiltonian in this region for the case where half quantum magnetic flux is penetrating the interior of the wire  
\begin{equation}
h_{B} (\theta) = \frac{1}{2 a_0} + \frac{1}{a_0} is_2 \partial_{\theta}
\end{equation}
To obtain the Hamiltonian in the region $C$ one has to start from Eq. (\ref{h01}) again since $R(z)$ is not well defined in this region. Using polar coordinates $(\rho, \varphi)$ the Hamiltonian takes a simple form and one can use the same unitary transformation given in Eq. (\ref{U01}) to put it in a rotationally invariant form. For the case where the half-quantum magnetic flux is uniformly distributed over the disk surface the $l=0$ Hamiltonian becomes  
\begin{equation}
h_C (\rho) = - i s_2 \partial_{\rho} -s_1 \frac{1}{2\rho} (1 - \frac{\rho^2}{R_0^2}) + s_3 m_{\rm Zeeman}
\end{equation}

Now since we assume that the SC order parameter vanishes in the $C$ region the zero-energy solutions to the BdG Hamiltonian in this region are the same as solutions of the above Hamiltonian. Assuming furthermore that the Zeeman field is negligible, we obtain a first order differential equation for each component of the spinor which is easy to solve and it leads to the solution of the form  given in Eq. (\ref{psiC}) 

For regions $A$ and $B$ the presence of the SC gap leads to four coupled linear differential equations which can be decoupled by a linear unitary transformation. What remains is to match the solutions at the two boundaries $A/B$ and $B/C$ and the result of this straightforward but somewhat tedious procedure is presented in Eqs. (\ref{psiA}-\ref{psiC}).



\begin{thebibliography}{1}
\bibitem{majorana} E. Majorana, Nuovo Cimento {\bf 5}, 171 (1937).
\bibitem{wilczek1} F. Wilczek, Nature Physics {\bf 5}, 614 (2009).
\bibitem{franz1} M. Franz, Physics {\bf 3}, 24 (2010).
\bibitem{alicea} J. Alicea, Rep. Prog. Phys. {\bf 75} 076501 (2012).
\bibitem{majreview} C. W. J. Beenakker, arXiv:1112.1950v2 (unpublished).
\bibitem{a12} C. Nayak, S. H. Simon, A. Stern, M. Freedman, and S. Das Sarma, Rev. Mod. Phys., 5{\bf 80}, 1083 (2008).
\bibitem{a42} M. H. Freedman, Proc. Natl. Acad. Sci., {\bf 95}, 98 (1998).
\bibitem{kitaev2} A. Kitaev,  Ann. Phys. {\bf 303}, 2 (2003).
\bibitem{a44} M. H. Freedman, A. Kitaev, M. J. Larsen, and Z. Wang, Bull. Amer. Math. Soc., {\bf 40}, 31 (2003).
\bibitem{a45} S. Das Sarma, M. Freedman, and C. Nayak, Phys. Rev. Lett., {\bf 94}, 166802 (2005).
\bibitem{a46} P. Bonderson, M. Freedman, and C. Nayak, Phys. Rev. Lett., {\bf 101}, 010501 (2008).
\bibitem{a49} L. Fu and C. L. Kane, Phys. Rev. Lett., {\bf 100}, 096407 (2008).
\bibitem{sau2}  R.M. Lutchyn, J.D. Sau and S. Das Sarma, \prl {\bf 105}, 077001 (2010).
\bibitem{oreg1} Y. Oreg, G. Refael and F. von Oppen, \prl {\bf 105}, 177002 (2010).
\bibitem{be150} V. Mourik, K. Zuo, S. M. Frolov, S. R. Plissard, E. P. A. M. Bakkers, and L. P. Kouwenhoven,  Science (April 12, 2012). 
\bibitem{be142} J. R. Williams, A. J. Bestwick, P. Gallagher, S. S. Hong, Y. Cui, A. S. Bleich, J. G. Analytis, I. R. Fisher, and D. Goldhaber-Gordon, \prl {\bf109} 056803 (2012)
\bibitem{furdyna1} L.P. Rokhinson, X. Liu and J. Furdyna, Nature Physics (2012).
\bibitem{cook} A. Cook and M. Franz, Phys. Rev. B {\bf 84}, 201105(R) (2011).
\bibitem{kongnanoribbon} D. Kong, {\em et al.}, Nano Lett. {\bf 10}, 329 (2010).
\bibitem{pengnanoribbon} H. L. Peng, {\em et al.}, Nature Mat. {\bf 9}, 225 (2010).
\bibitem{tangnanoribbon} H. Tang, {\em et al.}, ACS Nano {\bf 5}, 7510 (2011).
\bibitem{zhangnanoribbon} D. Zhang, {\em et al.}, Phys. Rev. B {\bf 84}, 165120 (2011)
\bibitem{alicea77} J. Alicea, Y. Oreg, G. Refael, F. von Oppen, M.P.A. Fisher, Nature Phys. {\bf 7}, 412 (2011). 
\bibitem{kitaev1} A.Y. Kitaev, Phys. Usp. {\bf 44}, 131 (2001).
\bibitem{nayak1} C. Nayak {\em et al.}, Rev. Mod. Phys. {\bf 80}, 1083 (2008).
\bibitem{shor} P. W. Shor, Phys. Rev. A 52, R2493 (1995).
\bibitem{or2} G. Moore and N. Read, Nucl. Phys. B 360, 362 (1991).
\bibitem{a8} G. E. Volovik, The Universe in a Helium Droplet (Oxford University Press, 2003).
\bibitem{a94} A. C. Potter and P. A. Lee, Phys. Rev. B, {\bf 83}, 184520 (2011).
\bibitem{a105} P. Anderson, Journal of Physics and Chemistry of Solids, {\bf 11}, 26 (1959).
\bibitem{a89} J. D. Sau, R. M. Lutchyn, S. Tewari, and S. Das Sarma, Phys. Rev. B, {\bf 82}, 094522 (2010).
\bibitem{a84} C. L. Kane and E. J. Mele, Phys. Rev. Lett., {\bf 95}, 226801 (2005).
\bibitem{a104} D. Xiao, W. Zhu, Y. Ran, N. Nagaosa, and So. Okamoto, Nat. Comm. {\bf 2}, 596 (2011).
\bibitem{a106} S. Murakami, Phys. Rev. Lett., {\bf 97}, 236805 (2006).
\bibitem{a107} B. A. Bernevig, T. L. Hughes, and S.-C. Zhang, Science, {\bf 314}, 1757 (2006).
\bibitem{a108} C. Liu, T. L. Hughes, X.-L. Qi, K. Wang, and S.-C. Zhang, Phys. Rev. Lett., {\bf 100}, 236601 (2008).
\bibitem{a109} C.-C. Liu, W. Feng, and Y. Yao, Phys. Rev. Lett., {\bf 107}, 076802 (2011).
\bibitem{a110} C. Weeks, J. Hu, J. Alicea, M. Franz, and R. Wu, Phys. Rev. X {\bf1}, 021001 (2011).
\bibitem{a111} M. Konig, S. Wiedmann, C. Brune, A. Roth, H. Buhmann, L. W. Molenkamp, X.-L. Qi, and S.-C. Zhang, Science, {\bf 325}, 766 (2007).
\bibitem{a112} A. Roth, C. Brune, H. Buhmann, L. W. Molenkamp, J. Maciejko, X.-L. Qi, and S.-C. Zhang, Science, {\bf 325}, 294 (2009).
\bibitem{a113} I. Knez, R.-R. Du, and G. Sullivan, Phys. Rev. Lett., {\bf 107}, 136603 (2011).
\bibitem{a114} G. Montambaux, Eur. Phys. J. B 79, 215 (2011).
\bibitem{a206} J. B. Miller, D. M. Zumb\"{u}hl, C. M. Marcus, Y. B. Lyanda-Geller, D. Goldhaber-Gordon, K. Campman, and A. C. Gossard, Phys. Rev. Lett., {\bf 90}, 076807 (2003).
\bibitem{a207} L. Meier, G. Salis, I. Shorubalko, E. Gini, S. Sch\"{o}n, and K. Ensslin, Nature Physics, {\bf 3}, 650 (2007).
\bibitem{a178} B. Seradjeh and E. Grosfeld, Phys. Rev. B, {\bf 83}, 174521 (2011).
\bibitem{a97} J. D. Sau, S. Tewari, and S. Das Sarma, Phys. Rev. B {\bf 85}, 064512 (2012).
\bibitem{a119} K. Flensberg, Phys. Rev. B, {\bf 82},  180516 (2010).
\bibitem{a120} A. R. Akhmerov, J. P. Dahlhaus, F. Hassler, M. Wimmer, and C. W. J. Beenakker, Phys. Rev. Lett., {\bf 106}, 057001 (2011).
\bibitem{a121} I. C. Fulga, F. Hassler, A. R. Akhmerov, and C. W. J. Beenakker, Phys. Rev. B, {\bf 83}, 155429 (2011).
\bibitem{a122} P. W. Brouwer, M. Duckheim, A. Romito, and F. von Oppen, Phys. Rev. B, {\bf 84}, 144526 (2011).
\bibitem{a123} P. W. Brouwer, M. Duckheim, A. Romito, and F. von Oppen, Phys. Rev. Lett., {\bf 107}, 196804 (2011).
\bibitem{a124} T. D. Stanescu, R. M. Lutchyn, and S. Das Sarma, Phys. Rev. B, {\bf 84}, 144522 (2011).
\bibitem{a125} D. Sticlet, C. Bena, and P. Simon, Phys. Rev. Lett. {\bf 108}, 096802 (2012).
\bibitem{a131} S. Nadj-Perge, V. S. Pribiag, J. W. G. van den Berg, K. Zuo, S. R. Plissard, E. P. A. M. Bakkers, S. M. Frolov, and L. P. Kouwenhoven, Phys. Rev. Lett. {\bf 108}, 166801 (2012).
\bibitem{a99} S. Gangadharaiah, B. Braunecker, P. Simon, and D. Loss, Phys. Rev. Lett., {\bf 107}, 036801 (2011).
\bibitem{a133} B. Braunecker, P. Simon, and D. Loss, Phys. Rev. B, {\bf 80}, 165119 (2009).
\bibitem{a134} E. M. Stoudenmire, J. Alicea, O. A. Starykh, and M. P. A. Fisher, Phys. Rev. B., {\bf 84}, 014503 (2011).
\bibitem{a103} A. Romito, J. Alicea, G. Refael, and F. von Oppen, Phys. Rev. B 85, 020502(R) (2012).
\bibitem{a135} M. Wimmer, A. R. Akhmerov, M. V. Medvedyeva, J. Tworzyd\l o, and C. W. J. Beenakker, Phys. Rev. Lett., {\bf 105}, 046803 (2010).
\bibitem{a143} M. Gibertini, F. Taddei, M. Polini, and R. Fazio, Phys. Rev. B {\bf 85}, 144525 (2012).
\bibitem{a146} J. D. Sau, C. H. Lin, H.-Y. Hui, and S. Das Sarma, Phys. Rev. Lett. {\bf 108}, 067001 (2012).
\bibitem{a147} W. DeGottardi, D. Sen, and S. Vishveshwara, New Yournal of Physics, {\bf 13}, 065028 (2011).
\bibitem{a148} J. D. Sau and S. Das Sarma, Nat. Comm. {\bf 3}, 964 (2012)
\bibitem{a151} J. Klinovaja, M. J. Schmidt, B. Braunecker, and D. Loss, Phys. Rev. B, {\bf 84}, 085452 (2011).
\bibitem{a152} J. D. Sau and S. Tewari, arXiv:1111.5622 (unpublished).
\bibitem{a153} J. Klinovaja, S. Gangadharaiah, and D. Loss, \prl {\bf 108}, 196804 (2012).
\bibitem{a154} L. P. Gor'kov and E. I. Rashba, Phys. Rev. Lett., {\bf 87}, 037004 (2001).
\bibitem{a155} M. Duckheim and P. W. Brouwer, Phys. Rev. B, {\bf 83}, 054513 (2011).
\bibitem{be138} A. Yu. Kasumov, O. V. Kononenko, V. N. Matveev, T. B. Borsenko, V. A. Tulin, E. E. Vdovin, and I. I. Khodos, Phys. Rev. Lett. {\bf 77}, 3029 (1996).
\bibitem{be139}D. Zhang, J. Wang, A. M. DaSilva, J. S. Lee, H. R. Gutierrez, M. H. W. Chan, J. Jain, and N. Samarth, Phys. Rev. B {\bf 84}, 165120 (2011).
\bibitem{be140} J. Wang, C.-Z. Chang, H. Li, K. He, D. Zhang, M. Singh, X.-C. Ma, N. Samarth, M. Xie, Q.-K. Xue, and M. H. W. Chan, Phys. Rev. B {\bf 85}, 045415 (2012).
\bibitem{be141} B. Sacepe, J. B. Oostinga, J. L. Li, A. Ubaldini, N. J. G. Couto, E. Giannini, and A. F. Morpurgo, Nat. Comm. 2, 575 (2011).
\bibitem{be143} M. Veldhorst, C. G. Molenaar, X. L. Wang, H. Hilgenkamp, and A. Brinkman, Appl. Phys. Lett. 100, 072602 (2012).
\bibitem{be144} F. Qu, F. Yang, J. Shen, Y. Ding, J. Chen, Z. Ji, G. Liu, J. Fan, X. Jing, C. Yang, and L. Lu, Scientific Reports {\bf 2}, 339 (2012).
\bibitem{be145} S. Sasaki, M. Kriener, K. Segawa, K. Yada, Y. Tanaka, M. Sato, and Y. Ando, Phys. Rev. Lett. {\bf 107}, 217001 (2011).
\bibitem{be146} G. Koren, T. Kirzhner, E. Lahoud, K. B. Chashka, and A. Kanigel, Phys. Rev. B {\bf 84}, 224521 (2011).
\bibitem{be147} F. Yang, Y. Ding, F. Qu, J. Shen, J. Chen, Z. Wei, Z. Ji, G. Liu, J. Fan, C. Yang, T. Xiang, and L. Lu, Phys. Rev. B {\bf 85}, 104508 (2012).
\bibitem{be148} M.-X. Wang, C. Liu, J.-P. Xu, F. Yang, L. Miao, M.-Y. Yao, C. L. Gao, C. Shen, X. Ma, X. Chen, Z.-A. Xu, Y. Liu, S.-C. Zhang, D. Qian, J.-F. Jia, and Q.-K. Xue, Science 336, 52Ð55 (2012).
\bibitem{plee_sept} J. Liu, A. C. Potter, K.T. Law, P. A. Lee, arXiv: 1206.1276.
\bibitem{tewari_sept} T. D. Stanescu and S. Tewari, arXiv:1208.6298v1.
\bibitem{ig1} M. T. Deng, C. L. Yu, G. Y. Huang, M. Larsson, P. Caroff, H. Q. Xu, arXiv:1204.4130.
\bibitem{ig2} A. Das, Y. Ronen, Y. Most, Y. Oreg, M. Heiblum, H. Shtrikman, arXiv:1205.7073.
\bibitem{mirlin1} P.~M.~Ostrovsky, I.~V.~Gornyi, A.~D.~Mirlin, \prl  {\bf 105}, 036803 (2010).
\bibitem{mele1} F. Zhang, C.L. Kane, and E.J. Mele, \prb {\bf 86}, 081303(R) (2012).
\bibitem{jackiw1} R. Jackiw and P. Rossi, Nucl. Phys B {\bf 190}, 681
  (1981).
\bibitem{qi_rev} X.-L. Qi, S.-C. Zhang, Rev. Mod. Phys. {\bf 83}, 1057-1110 (2011).
\bibitem{returnminigap} S. Tewari, T. D. Stanescu, J. D. Sau, S. Das Sarma, arXiv:1204.3637v2 (unpublished).
\bibitem{fu-berg1} L. Fu and E. Berg, \prl {\bf 105} 097001, (2010).
\bibitem{tinkham1} M. Tinkham, \textit{Introduction to Superconductivity} (McGraw Hill, New York) 1996.
\bibitem{neto77} N.M.R. Peres, F. Guinea, and A.H. Castro Neto, \prb {\bf 73}, 125411 (2006).
\bibitem{schnyder33} Andreas P. Schnyder, Shinsei Ryu, Akira Furusaki, Andreas W. W. Ludwig, Phys. Rev. B 78, 195125 (2008).
\bibitem{kitaev33} Alexei Kitaev,  arXiv:0901.2686.


\end{thebibliography}
\end{document}